\documentclass[aps, pre, a4paper, floatfix,  twocolumn, longbibliography, nofootinbib]{revtex4-1}

\usepackage[utf8]{inputenc}
\usepackage{epsfig}
\usepackage[T1]{fontenc}
\usepackage[english]{babel}
\usepackage[table]{xcolor}
\usepackage{t1enc}
\usepackage{graphicx}
\usepackage{amssymb}
\usepackage{amsmath}
\usepackage{relsize}
\usepackage[percent]{overpic}
\usepackage{bm}
\usepackage{hyperref}

\usepackage{siunitx}
\usepackage{accents}
\usepackage{mathtools}

\usepackage{grffile}

\usepackage[normalem]{ulem}

\usepackage{pgf}
\usepackage{tikz}

\newcommand{\avg}[1]{{\left<#1\right>}}

\newcommand{\dd}{\mathrm{d}}
\newcommand{\ee}{\mathrm{e}}

\usepackage{mathtools}
\def\multiset#1#2{\ensuremath{\left(\kern-.3em\left(\genfrac{}{}{0pt}{}{#1}{#2}\right)\kern-.3em\right)}}

\usepackage{amsmath}

\newcommand{\A}{\bm{A}}

\newcommand{\bb}{\bm{b}}

\usepackage{verbatim}
\usepackage{overpic}
\usepackage{booktabs}
\usepackage{placeins}

\usepackage{siunitx}
\usepackage{multirow}

\begin{document}

\title{Statistical inference of assortative community structures}

\author{Lizhi Zhang}
\email{lz848@bath.ac.uk}
\affiliation{Department of Mathematical Sciences, University of Bath, Claverton Down, Bath BA2
  7AY, United Kingdom}

\author{Tiago P. Peixoto}
\email{peixotot@ceu.edu}
\affiliation{Department of Network and Data Science, Central European University, H-1051 Budapest, Hungary}
\affiliation{ISI Foundation, Via Chisola 5, 10126 Torino, Italy}
\affiliation{Department of Mathematical Sciences, University of Bath, Claverton Down, Bath BA2
  7AY, United Kingdom}

\begin{abstract}
  We develop a principled methodology to infer assortative communities
  in networks based on a nonparametric Bayesian formulation of the
  planted partition model. We show that this approach succeeds in
  finding statistically significant assortative modules in networks,
  unlike alternatives such as modularity maximization, which
  systematically overfits both in artificial as well as in empirical
  examples. In addition, we show that our method is not subject to a
  resolution limit, and can uncover an arbitrarily large number of
  communities, as long as there is statistical evidence for them.  Our
  formulation is amenable to model selection procedures, which allow us
  to compare it to more general approaches based on the stochastic block
  model, and in this way reveal whether assortativity is in fact the
  dominating large-scale mixing pattern. We perform this comparison with
  several empirical networks, and identify numerous cases where the
  network's assortativity is exaggerated by traditional community
  detection methods, and we show how a more faithful degree of
  assortativity can be identified.
\end{abstract}

\maketitle
%\tableofcontents

\section{Introduction}

Community detection is one of the most central methods in network
science~\cite{fortunato_community_2010,fortunato_community_2016}, and it
consists in the algorithmic partition of the nodes of a network into
cohesive groups, according to a mathematical definition of this concept
(for which there are many). Historically, most community detection
methods proposed have focused on the detection of \emph{assortative}
communities,
i.e. groups of nodes that tend to be more connected to themselves than
to other nodes in the network. However, there are also community
detection methods that are more general, and attempt to cluster together
nodes that have similar patterns of connection, regardless if they are
assortative or
not~\cite{wasserman_stochastic_1987,reichardt_role_2007,karrer_stochastic_2011}.
The widespread use of assortative community detection methods has lead
to the belief that the presence of communities is a pervasive feature of
many different kinds of real
networks~\cite{newman_communities_2012}. Although the concept of
assortativity is a central one in the study of social networks (known as
``homophily'' in that context)~\cite{porter_communities_2009}, and is
also an appealing construct in
biology~\cite{guimera_functional_2005,ravasz_hierarchical_2002,holme_subnetwork_2003},
it is to some extent unclear if the perceived assortativity of many
networks is a byproduct of using algorithms that can only find this kind
of structure. This is particularly problematic since many popular
methods do not take into account the statistical significance of the
patterns they uncover, and find seemingly strong community structure in
completely random
graphs~\cite{guimera_modularity_2004,mcdiarmid_modularity_2016}, as well
as in trees~\cite{bagrow_communities_2012} and other manifestly
non-modular networks~\cite{mcdiarmid_modularity_2013}. More recently,
these shortcomings have been addressed by employing Bayesian inference
of generative network models~\cite{peixoto_bayesian_2019}, which
accounts for statistical significance with a built-in Occam's razor,
that decides to partition the network into groups only if this is
necessary to explain its structure, beyond what can be done by a
uniformly random placement of the edges. These approaches, however, are
based on general mixing patterns, which include assortativity only as a
special case. In many ways this is useful, and in fact arguably
superior, since if assortativity happens to be the dominating pattern,
then the general approach will capture it, otherwise it will reveal a
different structure. However, having only a more general method at our
disposal also has its shortcomings. First, if it is true that
assortativity is the main pattern for a class of networks, then the more
general representation is needlessly wasteful for them, since it not
only gives us more than we need, but in doing so it prevents us from
focusing on the more central features, at the cost of algorithmic
precision. Second, with a more general method it can be difficult to
quantify precisely how much has been wasted in the representation, and
what is indeed the simpler pattern hiding inside it.

In this work we develop a Bayesian inference approach designed to
uncover assortative communities in networks, based on the planted
partition (PP)
model~\cite{bui_graph_1987,dyer_solution_1989,condon_algorithms_2001},
which is itself a special case of the more general stochastic block
model (SBM)~\cite{wasserman_stochastic_1987,karrer_stochastic_2011}. Our
approach is nonparametric, and can uncover communities even when their
number is unknown, without overfitting. Furthermore we show that it does
not suffer from the resolution limit present in other approaches, such
as modularity maximization~\cite{fortunato_resolution_2007}, and it can
find an arbitrarily large number of communities, provided they are
statistically significantly. We also revisit an existing equivalence
between the inference of the PP model and modularity
maximization~\cite{zhang_scalable_2014,newman_equivalence_2016}, and
dispel the notion that both methods are interchangeable in practice, by
showing that the equivalence is in general inconsistent with maximum
likelihood estimation, and discuss the fact that even if this were not
the case, the parametric nature of that approach would not address the
overfitting problem. Our approach is not more complicated to employ than
modularity maximization, and can be used as a drop-in replacement for it
and other quality functions in popular community detection
heuristics~\cite{blondel_fast_2008}, and we describe how it can be used
with an unbiased merge-split MCMC
algorithm~\cite{peixoto_merge-split_2020} that can explore the entire
posterior distribution of partitions.

Furthermore, we perform a comparison of the PP model with the more
general SBM for a variety of empirical networks, allowing us to
determine if and to what extent is assortativity the most salient
characteristic of the large-scale network structure. We find a variety
of outcomes, ranging from very similar to very different results
obtained with both models, demonstrating that there are indeed many
cases where searching exclusively for assortative structures can give a
very misleading representation of the network.

This work is organized as follows. In Sec.~\ref{sec:pp} we present the
planted partition model, and we revisit its equivalence with
modularity. In Sec.~\ref{sec:bayes} we describe our Bayesian approach at
inferring the PP model, and introduce a more realistic non-uniform
version of the model. In Sec.~\ref{sec:artificial} we analyse the method
when applied to artificial networks with known community structures, and
compare it to variations of modularity-based approaches, demonstrating
that our method prevents overfitting. We also show that the Bayesian
method does not suffer from the resolution limit of modularity
maximization, and hence it does not generically underfit as well. In
Sec.~\ref{sec:empirical} we employ our method in a variety of empirical
networks, and we compare the results with those obtained by the more
general SBM, as well as modularity maximization. We demonstrate once
more that modularity tends to either massively overfit or underfit, and
when comparing to the SBM we can determine the true nature of
the assortativity of empirical networks. We finalize in
Sec.~\ref{sec:conclusion} with a conclusion.

\section{The Planted Partition (PP) model}\label{sec:pp}

The statistical inference approach to community detection is based on
the definition of generative models that contain communities as part of
the their parameters. Before we consider the particular case of
inferring assortative communities, it is useful at first to review the
more general case of arbitrary mixing patterns between groups, as
characterized by the Poisson degree-corrected stochastic block model
(DC-SBM)~\cite{karrer_stochastic_2011}. This formulation describes a
network of $N$ nodes that are divided into $B$ groups, amounting to a
partition $\bb=\{b_i\}$, where $b_i \in [1,B]$ is the group membership
of node $i$, and a specific multigraph $\A$ is generated with
probability
\begin{multline}\label{eq:sbm}
    P(\A|\bm\lambda, \bm\theta, \bb) = \prod_{i<j} \ee^{-\theta_i\theta_j\lambda_{b_ib_j}}\frac{(\theta_i\theta_j\lambda_{b_ib_j})^{A_{ij}}}{A_{ij}!}\\\times \prod_i \ee^{-\theta_i^2\lambda_{b_ib_i}/2}\frac{(\theta_i^2\lambda_{b_ib_i}/2)^{A_{ii}/2}}{(A_{ii}/2)!!},
\end{multline}
where $A_{ij}$ determines the number of edges between nodes $i$ and $j$,
and by convention $A_{ii}$ corresponds to twice the number of self-loops
incident on node $i$. Note that the parameters $\bm\lambda$ and
$\bm\theta$ always appear multiplying each other in the likelihood, so
they are not uniquely identifiable, i.e. there are many choices that
yield the same model. In order to uniquely specify the model, it is
useful to introduce the quantity
\begin{equation}
  \hat\theta_r = \sum_i \theta_i\delta_{b_ir},
\end{equation}
which in the model above can be set to arbitrary values, without
sacrificing its generality. For example, if we set $\hat\theta_r=1$,
then we can interpret $\lambda_{rs}$ as determining the expected number
of edges between groups $r$ and $s$ (or twice this value for $r=s$), and
$\theta_i$ is the relative probability with which a node $i$ is selected
to form an edge among those that belong to the same group. However, any
other choice for $\hat\theta_r$ would be equally valid, with the only
immaterial consequence being a different interpretation of the
parameters. The maximum likelihood estimate of the above parameters is
given by
\begin{equation}
  \lambda^*_{rs} = \frac{e_{rs}}{\hat\theta_r\hat\theta_s}, \quad
  \theta_i^* = \frac{k_i}{e_{b_i}}\hat\theta_{b_i},
\end{equation}
where $e_{rs}$ is the number of edges that go between groups $r$ and $s$
(or twice that for $r=s$), and
$e_r=\sum_se_{rs}=\sum_ik_i\delta_{b_i,r}$ is the sum of degrees in
group $r$.  Indeed, the values of $\hat\theta_r$ cannot be uniquely
obtained from these equations, since any value $\hat\theta_r>0$ offers a
valid solution, and more importantly, any choice disappears when we
compute the probabilities $\lambda_{b_ib_j}^*\theta_i^*\theta_j^*$. This
means we can choose these values independently of the inference
procedure, with any particular choice functioning as a mere technical
convention.\footnote{Strictly speaking, this is no longer true in a
Bayesian setting, where we are required to integrate the likelihood over
the set of parameters. As shown in
Ref.~\cite{peixoto_nonparametric_2017}, choosing $\hat\theta_r$ has an
effect on the parameter space and choice of priors, and ultimately
changes the integrated marginal likelihood.}

The degree-corrected planted partition (PP) model corresponds to the
special case of the DC-SBM given by
\begin{equation}
  \lambda_{rs} = \lambda_{\text{in}} \delta_{rs} + \lambda_{\text{out}}(1 - \delta_{rs}).
\end{equation}
In this situation there are only two parameters that determine the
placement of edges between groups, $\lambda_{\text{in}}$ and
$\lambda_{\text{out}}$, that set the expected number of edges inside and
outside groups. A choice $\lambda_{\text{in}}\sum_r\hat\theta_r^2/2 >
\lambda_{\text{out}}\sum_{r<s}\hat\theta_r\hat\theta_s$ corresponds to
the assortative case, where edges connect mostly nodes of the same
group. Therefore, this model captures what is more typically known as
community structure in the proper sense, at least if the condition just
mentioned is met. With this parametrization, the model likelihood of
Eq.~\ref{eq:sbm} becomes
\begin{multline}\label{eq:pp_likelihood}
  P(\A|\lambda_{\text{in}},\lambda_{\text{out}}, \bm\theta, \bb)
  = \\
  \frac{\ee^{-\lambda_{\text{out}}\sum_{r<s}\hat\theta_r\hat\theta_s}\lambda_{\text{out}}^{e_{\text{out}}}\ee^{-\lambda_{\text{in}}\sum_r\hat\theta_r^2/2}\lambda_{\text{in}}^{e_{\text{in}}}\prod_i \theta_i^{k_i}}{\prod_{i<j}A_{ij}!\prod_i A_{ii}!!},
\end{multline}
where,
\begin{equation}
  e_{\text{in}} = \frac{1}{2}\sum_{ij}A_{ij}\delta_{b_i,b_j},\quad
  e_{\text{out}} = \sum_{i<j}A_{ij}(1-\delta_{b_i,b_j}),
\end{equation}
are the number of edges inside and outside groups, respectively.  The
maximum likelihood estimate of the parameters of the PP model are then
given by
\begin{align}
  \lambda_{\text{in}}^* &= \frac{2e_{\text{in}}}{\sum_r\hat\theta_r^2}\label{eq:mle_lin}\\
  \lambda_{\text{out}}^* &= \frac{e_{\text{out}}}{\sum_{r<s}\hat\theta_r\hat\theta_s}\label{eq:mle_lout}\\
  \theta_i^* &= k_i \left[\frac{2e_{\text{in}}\hat\theta_{b_i}}{\sum_r\hat\theta_r^2}+\frac{e_{\text{out}}\sum_{r\neq b_i}\hat\theta_r}{\sum_{r<s}\hat\theta_r\hat\theta_s}\right]^{-1}.\label{eq:mle_theta}
\end{align}
Looking at this result, we see that, unlike in the general DC-SBM, we no
longer have full freedom to choose $\hat\theta_r$, since its maximum
likelihood value must be a solution of the following system of nonlinear
equations
\begin{equation}
  \hat \theta_r^* = e_r \left[\frac{2e_{\text{in}}\hat\theta_r^*}{\sum_s\hat{\theta_s^*}^2}+\frac{e_{\text{out}}\sum_{s\neq r}\hat\theta_s^*}{\sum_{s<t}\hat\theta_s^*\hat\theta_t^*}\right]^{-1}.
\end{equation}
We recover partial freedom to choose $\hat\theta_r$ in the special
situation where all groups are uniform, with $e_r=2E/B$, in which case
$\hat\theta_r$ can take any value, as long as it is the same for every
group, i.e. $\hat\theta_r=\hat\theta$. Note that we are, in fact,
allowed to make an arbitrary prior assumption for the values of
$\hat\theta_r$ before doing inference, making them imposed constraints
that are part of our model specification. In this case
Eq.~\ref{eq:mle_theta} becomes simply
\begin{equation}
  \theta_i^* = \frac{k_i\hat\theta_r}{e_{b_i}}\label{eq:mle_theta_alt}.
\end{equation}
We emphasize, however, that imposing this sort of constraint does
jeopardize the degree correction of the model, since the expected degree
of node $i$ is given by
\begin{align}
  \avg{k_i}
  &= \sum_j \theta_i\theta_j[\lambda_{\text{in}} \delta_{b_i,b_j} + \lambda_{\text{out}}(1 - \delta_{b_i,b_j})]\\
  &= \theta_i\left[\hat\theta_{b_i}\lambda_{\text{in}} + \lambda_{\text{out}}\sum_{r\ne b_i}\hat\theta_r\right].
\end{align}
If we now substitute the maximum likelihood estimates of
Eqs.~\ref{eq:mle_theta_alt}, \ref{eq:mle_lin} and~\ref{eq:mle_lout} in
the above, we obtain
\begin{align}
  \avg{k_i} = \frac{k_i\hat\theta_{b_i}}{e_r} \left[\frac{2e_{\text{in}}\hat\theta_{b_i}}{\sum_r\hat\theta_r^2}+\frac{e_{\text{out}}\sum_{r\neq b_i}\hat\theta_r}{\sum_{r<s}\hat\theta_r\hat\theta_s}\right].
\end{align}
Therefore, the inferred model generates the observed degrees in
expectation, i.e. $\avg{k_i}=k_i$, only if all groups have the same sum
of degrees $e_r=2E/B$ and all imposed $\hat\theta_r$ are the same, or if
we do not impose any constraints on $\hat\theta_r$, and use
Eq.~\ref{eq:mle_theta} instead. This means that we face a trade-off
between consistent degree correction and ease of inference with the
planted partition model, which is important to keep in mind as we
consider the connection between statistical inference and modularity
maximization, which we address in the following.

\subsection{On the consistency between statistical inference and modularity maximization}

As was shown in
Refs.~\cite{zhang_scalable_2014,newman_equivalence_2016}, it is possible
to manipulate the likelihood of the PP model to expose a connection with
modularity maximization~\cite{newman_modularity_2006}. We can rewrite
the likelihood of Eq.~\ref{eq:pp_likelihood} as
\begin{multline}
  \ln P(\A|\lambda_{\text{in}},\lambda_{\text{out}}, \bm\theta, \bb)\\
  = \frac{\mu}{2} \sum_{ij}\left( A_{ij} - \gamma\theta_i\theta_j\right)\delta_{b_ib_j}+ E \ln \lambda_{\text{out}}\\
  - \frac{\lambda_{\text{out}}}{2}\left(\sum_i\theta_i\right)^2 + \sum_ik_i\ln\theta_i,
\end{multline}
up to unimportant additive constants, and where
\begin{equation}\label{eq:mu_gamma}
  \mu = \ln \lambda_{\text{in}} - \ln \lambda_{\text{out}}, \quad \gamma = \frac{\lambda_{\text{in}} - \lambda_{\text{out}}}{\ln \lambda_{\text{in}} - \ln \lambda_{\text{out}}}.
\end{equation}
If we now enforce the following constraint as part of our model
specification
\begin{equation}\label{eq:q_const}
  \hat\theta_r = \frac{e_r}{\sqrt{2E}},
\end{equation}
and replace the maximum likelihood estimate for
$\theta^*_i=k_i/\sqrt{2E}$ obtained from Eq.~\ref{eq:mle_theta_alt} in
the above we obtain
\begin{multline}\label{eq:pre_Q}
  \ln P(\A|\lambda_{\text{in}},\lambda_{\text{out}}, \bm\theta=\bm\theta^*, \bb)\\
  = \frac{\mu}{2} \sum_{ij}\left( A_{ij} - \gamma\frac{k_ik_j}{2E}\right)\delta_{b_ib_j} \\
  + E(\ln \lambda_{\text{out}} - \lambda_{\text{out}}),
\end{multline}
again up to unimportant additive constants. Therefore, maximizing the
above likelihood with respect to the partition $\bb$ alone, while
keeping $\lambda_{\text{in}}$ and $\lambda_{\text{out}}$ constant, is
equivalent to maximizing the generalized
modularity~\cite{reichardt_statistical_2006}
\begin{equation}\label{eq:Q}
  Q(\A,\bb) = \frac{1}{2E}\sum_{ij}\left(A_{ij} - \gamma \frac{k_ik_j}{2E}\right),
\end{equation}
with $\gamma$ playing the role of the resolution parameter. However,
before concluding that modularity maximization and the inference of the
PP model amount to the same task, we need to make the following crucial
observations:
\begin{enumerate}
\item The imposed constraint of Eq.\ref{eq:q_const} involves the
  knowledge of the sum of all \emph{observed} degrees in each group
  $e_r=\sum_ik_i\delta_{b_i,r}$, which cannot be known before doing
  inference, and thus cannot be part of our model
  specification.\footnote{The same is true for the total number of
    edges $E$, but to a lesser degree since this is a global value that
    does not depend on the network partition (and hence amounts to a
    more innocuous inconsistency), and replacing it by any other
    constant would only amount to a different effective value of
    $\gamma$ in Eq.~\ref{eq:Q}.} However, any other choice of
  $\hat\theta_r$ will not yield $\theta^*_i=k_i/\sqrt{2E}$ via maximum
  likelihood, which is required to recover modularity. Not imposing any
  prior constraint on $\hat\theta_r$ also does not yield the appropriate
  value via Eq.~\ref{eq:mle_theta} in general, and will result in the
  necessary value of $\theta^*_i$ only for a particular uniform
  partition of the network where $e_r=2E/B$ (in which case
  Eq.~\ref{eq:q_const} holds without being imposed). Therefore, the
  modularity of Eq.~\ref{eq:Q} is consistent with maximum likelihood of
  the PP model only in the very narrow case where all groups have the
  same sum of degrees.
\item In addition, we must keep in mind that the
  values of $\lambda_{\text{in}}$ and $\lambda_{\text{out}}$ are never
  known \emph{a priori} in empirically relevant settings. Therefore, we
  are required to infer them as well, together with the network
  partition. When employing maximum likelihood, the resulting values of
  $\mu$ and $\gamma$, as well as the second term of Eq.~\ref{eq:pre_Q}
  all depend on the network partition, and are no longer just
  constants. In this situation, the partial equivalence with modularity
  maximization breaks down (even if $e_r=2E/B$ as per point 1 above), as
  the functional form resulting from substituting
  Eqs.~\ref{eq:mle_lin}-\ref{eq:mle_theta} and Eq.~\ref{eq:mu_gamma}
  into Eq.~\ref{eq:pre_Q} makes the latter very different from
  Eq.~\ref{eq:Q}. We emphasize that the scheme suggested in
  Ref.~\cite{newman_equivalence_2016} of updating the value of $\gamma$
  according to Eq.~\ref{eq:mu_gamma} is insufficient to restore
  consistency since the contribution of the non-negligible terms $\mu$
  and $E(\ln \lambda_{\text{out}} - \lambda_{\text{out}})$ remain
  unaccounted for.
\end{enumerate}

Based on the above, we see that the overall connection between the
inference of the PP model and modularity maximization is in fact rather
tenuous, and we should not expect in general to obtain the same results
with both approaches. As explained in Ref.~\cite{zhang_scalable_2014},
the only statement that can be made is that there exists a particular
choice of parameters $\lambda_{\text{in}}$, $\lambda_{\text{out}}$ and
$\bm\theta$ such that maximizing modularity with the appropriate choice
of $\gamma$ and the PP likelihood conditioned on these parameters will
yield the same partition. But since these parameters are unknown in
practice, and are in general inconsistent with maximum likelihood
estimation, the relevance of this equivalence is arguably limited.

Furthermore, as we discuss in Appendix~\ref{app:equiv}, it is easy to
establish a formal equivalence between any community detection method
and the statistical inference of a suitably chosen generative
model. Therefore, the central issue is not whether this mapping exists,
but if the procedure itself is consistent and behaves well. In fact,
neither approach considered above, i.e. maximum likelihood inference of
the PP model and modularity maximization, actually offers a robust
method to uncover community structure in networks. As is well known,
modularity maximization suffers from severe shortcomings, such as a
strong tendency to identify spurious communities in fully
random~\cite{guimera_modularity_2004} and
non-modular~\cite{mcdiarmid_modularity_nodate,bagrow_communities_2012,mcdiarmid_modularity_2013,mcdiarmid_modularity_2016}
networks, a systematic failure to identify relatively small communities
in large networks~\cite{fortunato_resolution_2007}, it exhibits extreme
degeneracy in key empirically relevant
cases~\cite{good_performance_2010}, and has been recently shown to
systematically overfit on a broad range of empirical
networks~\cite{ghasemian_evaluating_2019}. Any equivalence with the
statistical inference of a parametric model would just mean that the
latter also inherits these same limitations. However, the full maximum
likelihood inference approach of the PP model outlined above (which is
not equivalent to modularity optimization) is not substantially
superior. Even though it has a better justification, it does not really
address any of core problems of modularity. Most prominently, the
inference approach is still prone to overfitting, with the uncontrolled
detection of an ever increasing number of meaningless communities in
fully random networks, as long as those increase the likelihood of the
model. This happens in the same manner as fitting a polynomial to a set
of points will also overfit, even if we use maximum likelihood, as long
as we are allowed to increase the polynomial order without any
constraint. We will demonstrate this problem with some simple examples,
but before we do so we turn instead to a Bayesian approach, which
includes the correct penalization of model complexity, and hence
addresses the overfitting problem at its root, in a manner analogous to
what has been done for the general SBM~\cite{peixoto_bayesian_2019}, as
we describe in the next session.

\section{Bayesian inference of the planted partition model}\label{sec:bayes}

Instead of maximum likelihood, a more formally correct approach to
statistical inference is to sample or maximize from the posterior
distribution of partitions~\cite{peixoto_bayesian_2019}
\begin{equation}\label{eq:posterior}
  P(\bb|\A) = \frac{P(\A|\bb)P(\bb)}{P(\A)},
\end{equation}
where
\begin{multline}\label{eq:pp_marginal}
 P(\A|\bb) = \int P(\A|\lambda_{\text{in}},\lambda_{\text{out}},\bm\theta,\bb)\times\\P(\lambda_{\text{in}})P(\lambda_{\text{out}})P(\bm\theta|\bb)\,\dd\lambda_{\text{in}}\dd\lambda_{\text{out}}\dd\bm\theta
\end{multline}
is the marginal likelihood integrated over all model parameters,
weighted according to their prior probabilities, and
\begin{equation}\label{eq:bprior}
  P(\bb) = \frac{\prod_r n_r!}{N!}{N-1\choose B(\bb)-1}^{-1}\frac{1}{N},
\end{equation}
is the prior probability for partition $\bb$, with $B(\bb)$ denoting the
number of groups of $\bb$ (see Ref.~\cite{peixoto_nonparametric_2017}
for a derivation). The remaining term $P(\A)=\sum_{\bb}P(\A|\bb)P(\bb)$
is called the evidence, and it has the role of a normalization constant,
and therefore will not play an important role in our calculations.  In
order to compute the integral of Eq.~\ref{eq:pp_marginal} we must
specify our priors, which involves us also dealing with the model
specification problem exposed earlier, with respect to the parameters
$\bm\theta$. Here we will make the simple choice
\begin{equation}
  \hat\theta_r=1,
\end{equation}
which allows the model parameters to have a straightforward
interpretation, namely the $\theta_i$ are the relative probabilities of
selecting a node randomly from the group it belongs, and
$B\lambda_{\text{in}}$ will determine twice the expected total number of
edges inside communities, and ${B\choose 2}\lambda_{\text{in}}$ the
number of edges outside communities. (Remember that we are allowed to
make any choice of $\hat\theta_r$ as part of our model specification, as
long as the choice is made \emph{a priori}, and does not depend on the
data being modelled. As discussed previously, this choice does limit the
accuracy of degree correction of the PP model when performing maximum
likelihood estimation. However, as we will see in a moment, this will
not be a problem in the Bayesian formulation.) Now we can proceed in a
manner that reflects our \emph{a priori} indifference to any kind of
model pattern, namely we select a uniform prior for $\bm\theta$,
\begin{equation}
  P(\bm\theta|\bb)=\prod_r(n_r-1)!\delta\left(\textstyle\sum_i\theta_i\delta_{b_i,r}-1\right),
\end{equation}
and we choose maximum-entropy priors for the remaining parameters
\begin{align}
  P(\lambda_{\text{in}}|\bar\lambda)&=\ee^{-\lambda_{\text{in}}/(2\bar\lambda)}/(2\bar\lambda),\\
  P(\lambda_{\text{out}}|\bar\lambda)&=\ee^{-\lambda_{\text{out}}/\bar\lambda}/\bar\lambda,
\end{align}
where $\bar\lambda$ is a hyperparameter that determines the expected
total number of edges, with $\bar\lambda=2\avg{E}/B^2$. Performing the
integral of Eq.~\ref{eq:pp_marginal} we obtain
\begin{multline}\label{eq:pp_canonical}
  P(\A|\bar\lambda,\bb) = \frac{e_{\text{in}}!e_{\text{out}}!}
  {2\bar\lambda^2\left[\frac{B}{2}+\frac{1}{2\bar\lambda}\right]^{e_{\text{in}}+1}\left[{B\choose 2}+\frac{1}{\bar\lambda}\right]^{e_{\text{out}}+1}}
  \times \\
  \prod_r\frac{(n_r-1)!}{(e_r+n_r-1)!}\times\frac{\prod_ik_i!}{\prod_{i<j}A_{ij}!\prod_i A_{ii}!!}.
\end{multline}
This marginal likelihood still depends on global hyperparameter
$\bar\lambda$, which we can infer together with the other model
parameters. However, there is an alternative that allows us to remove it
altogether. We can re-interpret this marginal likelihood as an entirely
equivalent model formulation given by
\begin{multline}\label{eq:pp_micro}
  P(\A|\bar\lambda,\bb) = P(\A|\bm e, \bm k, \bm b)P(\bm k |\bm e,\bb)P(\bm e|e_{\text{in}},e_{\text{out}},\bb)\times \\
  P(e_{\text{in}}|\bar\lambda)P(e_{\text{out}}|\bar\lambda),
\end{multline}
where
\begin{equation}
  P(\A|\bm e, \bm k, \bm b) = \frac{\prod_{r<s}e_{rs}!\prod_re_{rr}!!\prod_ik_i!}{\prod_re_r!!\prod_{i<j}A_{ij}!\prod_iA_{ii}!}
\end{equation}
is the likelihood of the microcanonical
DC-SBM~\cite{peixoto_nonparametric_2017}, where $e_{rs}$ specifies the
exact number of edges between groups $r$ and $s$ (or twice that for
$r=s$) and $k_i$ is the exact degree of node $i$. We can recover
Eq.~\ref{eq:pp_canonical} by making the following choice of priors,
\begin{equation}
  P(\bm e|e_{\text{in}}, e_{\text{out}}, \bb) =
  \frac{e_{\text{in}}!}{B^{e_{\text{in}}}\prod_r(e_{rr}/2)!} \times
  \frac{e_{\text{out}}!}{{B\choose 2}^{e_{\text{out}}}\prod_{r<s}e_{rs}!},
\end{equation}
which is a product of uniform multinomial distributions for the diagonal
and off-diagonal entries of the matrix $e_{rs}$, conditioned on the
total sums $e_{\text{in}}$ and $e_{\text{out}}$, respectively.  For
$e_{\text{in}}$ and $e_{\text{out}}$ themselves we use geometric
distributions,
\begin{align}
  P(e_{\text{in}}|\bar\lambda,\bb) &= \frac{\left(B\bar\lambda\right)^{e_{in}}}{\left(B\bar\lambda+1\right)^{e_{in}+1}}\\
  P(e_{\text{out}}|\bar\lambda,\bb) &= \frac{\left({B\choose 2}\bar\lambda\right)^{e_{out}}}{\left({B\choose 2}\bar\lambda+1\right)^{e_{out}+1}},
\end{align}
and finally for the degrees we choose uniform distributions inside each
group~\cite{peixoto_nonparametric_2017},
\begin{equation}
  P(\bm k|\bm e,\bb) = \prod_r\frac{e_r!(n_r-1)!}{(e_r+n_r-1)!}.
\end{equation}
Inserting these priors in Eq.~\ref{eq:pp_micro} and re-arranging leads
to Eq.~\ref{eq:pp_canonical}. Interestingly, and somewhat surprisingly,
since this microcanonical model generates the exact degrees $\bm k$ that
are observed, we no longer have the same inconsistency as in the
``canonical'' model under maximum likelihood that we discussed earlier,
where the inferred degrees were different from the observed, even though
we have made use of the constraint $\hat\theta_r=1$ in its
derivation. We can therefore rest assured this model can accommodate
arbitrary degree sequences. This equivalence also allows us to replace
some of the priors of the microcanonical formulation by more convenient
choices that make the approach fully nonparametric. In particular, for
$e_{\text{in}}$ and $e_{\text{out}}$ we can use instead the following
\begin{equation}
  P(e_{\text{in}}, e_{\text{out}}|\bb) = P(e_{\text{in}}, e_{\text{out}}|E,\bb)P(E)
\end{equation}
where
\begin{equation}
  P(e_{\text{in}}, e_{\text{out}}|E,\bb) = \left(\frac{1}{E+1}\right)^{1-\delta_{B,1}}
\end{equation}
is a uniform distribution of the $E$ edges into two values (unless
$B=1$, where we must have $e_{\text{in}}=E$). The prior for the total
number of edges $P(E)$ can now be chosen arbitrarily, as it will only
amount to a unimportant constant in the marginal distribution, and hence
vanish from the posterior. With this, we have a fully non-parametric
marginal distribution
\begin{multline}
  P(\A|\bb) = P(\A|\bm e, \bm k, \bm b)P(\bm k |\bm e,\bb)\times\\
  P(\bm e|e_{\text{in}},e_{\text{out}}|\bb)P(e_{\text{in}}, e_{\text{out}}|E,\bb)P(E)
\end{multline}
that reads
\begin{multline}
  P(\A|\bb) =
  \frac{e_{\text{in}}!e_{\text{out}}!}
  {\left(\frac{B}{2}\right)^{e_{\text{in}}}{B\choose 2}^{e_{\text{out}}}(E+1)^{1-\delta_{B,1}}}\times\\
  \prod_r\frac{(n_r-1)!}{(e_r+n_r-1)!}\times\frac{\prod_ik_i!}{\prod_{i<j}A_{ij}!\prod_i A_{ii}!!}.
\end{multline}
This expression, together with the partition prior $P(\bb)$ of
Eq.~\ref{eq:bprior}, are not much more difficult to compute than the
modularity of Eq.~\ref{eq:Q}. In fact, it is easy to see that if we
consider the change in the posterior probability that is incurred if we
move a node $i$ from group $r$ to group $s$, we need to compute only a
few terms that depend on $e_{\text{in}}$, $e_{\text{out}}$, $e_r$,
$e_s$, $n_r$, $n_s$ and $B$. In order to compute the change, we need
only to inspect the neighborhood of the node, which takes time $O(k_i)$,
independently of any other quantity, such as the number of groups.  This
is the same algorithmic complexity of computing changes in modularity,
so the quantity $\ln P(\A,\bb)$ can be used a drop-in replacement of the
quality function in any modularity maximization algorithm,\footnote{For
reasons of numerical stability, it is better to work with the
log-probability $\ln P(\A,\bb)$, such that products become sums, and we
can also use tables of pre-computed log-factorial values to improve the
speed of the computation.} thereby addressing many existing fundamental
limitations. In fact, we can understand in more detail why this approach
prevents overfitting by exploiting a direct connection between Bayesian
inference and information theory. Namely we can write the negative joint
log-likelihood as follows,
\begin{align}
  \Sigma &= -\ln P(\A,\bb) = -\ln P(\A,\bm k,\bm e,\bb)\\
         &= -\ln P(\A|\bm e, \bm k, \bm b)-\ln P(\bm e, \bm k, \bm b).
\end{align}
The quantity $\Sigma$ is called the description length of the
data~\cite{grunwald_minimum_2007}, as it measures the amount of
information required to describe the network $\A$ when the parameters
$\bm e$, $\bm k$, and $\bm b$ are known, together with the information
necessary to describe the parameters themselves.  Therefore, the most
likely partition of the network is the one that allow us to compress it
the most. This means that this approach amounts to a formal
implementation of Occam's razor, that favors the most parsimonious
explanation for the data: As we increase the complexity of the model, by
considering a larger number of communities, the first term $-\ln
P(\A|\bm e, \bm k, \bm b)$ will tend to decrease, as the model becomes
more constrained, however the second term $-\ln P(\bm e, \bm k, \bm b)$
will tend to increase, functioning as a penalty for more complex
models. Since it is not possible to compress fully random data using any
method, this approach cannot find communities in fully random
networks. (This also explains why maximum likelihood overfits, since it
omits the contribution of the second term of the description length, and
hence there is no penalization for model complexity.) We will also show
in Sec.~\ref{sec:resolution} that this approach also does not suffer
from the ``resolution limit'' underfitting problem present with
modularity maximization. However, before we do so, we will first
consider a small variation of the PP model that is slightly more
realistic, and allows for a larger amount of heterogeneity in the
community structure.

\subsection{The non-uniform PP model}

Even if we commit ourselves to search exclusively for assortative
community structures, the particular formulation of the PP model
considered previously seems needlessly restrictive. This is because it
assumes that, if all $\theta_i$ are the same, then the expected number
of edges inside communities is the same for every community, which is
likely to be an inadequate assumption in a variety of empirical
scenarios. We can relax this constraint by formulating instead a
non-uniform version of the PP model, with
\begin{equation}
  \lambda_{rs} = \lambda_r \delta_{rs} + \omega(1 - \delta_{rs}).
\end{equation}
This parametrization allows for the expected number of edges inside
communities to vary arbitrarily, via the parameters
$\bm\lambda=\{\lambda_r\}$ that can be different for every
community. Given this formulation, we can essentially repeat the same
calculations as before, as we show in Appendix~\ref{app:nupp}. In the
end, we obtain a marginal likelihood given by
\begin{multline}\label{eq:nupp}
  P(\A|\bb) =
  \frac{e_{\text{out}}!\prod_re_{rr}!!}
       {{B\choose 2}^{e_{\text{out}}}(E+1)^{1-\delta_{B,1}}}\times{B + e_{\text{in}} - 1 \choose e_{\text{in}}}^{-1}\times\\
       \prod_r\frac{(n_r-1)!}{(e_r+n_r-1)!}\times\frac{\prod_ik_i!}{\prod_{i<j}A_{ij}!\prod_i A_{ii}!!}.
\end{multline}
This likelihood is very similar to the uniform planted partition model,
and is just as easy to compute, but it should work better when the
communities are sufficiently heterogeneous.

\subsection{Inference algorithm}

The posterior distribution of Eq.~\ref{eq:posterior} is not simple
enough to allow us to sample directly from it, so we have to perform
this indirectly using Markov chain Monte Carlo (MCMC). This is done by
defining move proposals that are conditioned on the current partition
$\bb$,
\begin{equation}
  P(\bb'|\bb)
\end{equation}
and accepting a new partition $\bb'$ sampled from this distribution
according to the Metropolis-Hastings
probability~\cite{metropolis_equation_1953,hastings_monte_1970}
\begin{equation}
  \min\left(1, \frac{P(\bb'|\A)P(\bb|\bb')}{P(\bb|\A)P(\bb'|\bb)}\right),
\end{equation}
otherwise we reject the move, and remain at the previous partition
$\bb$. Note that the computation of the above ratio does not depend on
the intractable normalization constant $P(\A)$ of
Eq.~\ref{eq:posterior}, since it cancels out in the computation. By
iterating the above procedure sufficiently often, we are guaranteed to
sample from the target distribution $P(\bb|\A)$ asymptotically, provided
our proposals $P(\bb'|\bb)$ are ergodic and aperiodic. However, the time
required to reach the target distribution will depend on the quality of
our proposals, which will determine the practical feasibility of the
algorithm. In this work we use the merge-split proposals described in
Ref.~\cite{peixoto_merge-split_2020}, which have been shown to work well
in many cases, in particular when the number of groups tends to
vary. The only modification we make of that algorithm is that when
proposing the move of a single node $i$ from its current group to group
$r$, we do it according to the following probability,
\begin{equation}
  P(r|i,\epsilon) = (1-\epsilon)\frac{\sum_jA_{ij}\delta_{b_j,r}}{k_i} + \frac{\epsilon}{B},
\end{equation}
where $B$ is the number of occupied groups. The parameter $\epsilon$
determines the probability with which we look at a random neighbor of
node $i$ to copy its group membership, otherwise we select a group at
random. We require a value $\epsilon>0$ to guarantee ergodicity, but
otherwise any other value yields a valid algorithm (we have used
$\epsilon=1/2$ in our analysis, which provided good acceptance rates).

As we mentioned before, when using our model formulation, the likelihood
ratio when changing the membership of node $i$, as well as the move
proposal probability, can be computed in time $O(k_i)$, where $k_i$ is
its degree. This means that a single MCMC ``sweep'', where every node
had a chance to be moved once, takes a time $O(N + E)$, where $E$ is the
number of edges, which is the best we can hope for this kind of
problem. Therefore we can use this algorithm to approach networks with a
very large size.  A reference implementation of this algorithm is freely
available as part of the \texttt{graph-tool}
library~\cite{peixoto_graph-tool_2014}.

In some cases, we may seek to maximize from the posterior distributions
instead of sampling from it. This is achieved via a simple modification
of the above algorithm, where we replace the target distribution with
$P(\bb|A)\to P(\bb|A)^{\beta}$, where $\beta$ is an inverse temperature
parameter. If we increase $\beta\to\infty$ (preferably slowly, to avoid
getting trapped in local optima) we obtain a maximization algorithm. The
merge-split MCMC often shows a good behavior when employed in this
manner, as it can more easily escape local maxima that would trap
alternative schemes, such as those based on the change of a single node
at a time.

\section{Artificial networks}\label{sec:artificial}

Here we show how our approach behaves for artificial networks that have
imposed community structure. We compare the inference of the PP model
with the DC-SBM~\cite{peixoto_nonparametric_2017}, as well as with
variations of modularity. We focus on the overfitting problem, and the
potential identification of non-existing communities. We do so by
sampling networks with $N=10^5$ nodes and average degree $\avg{k}=5$,
and a specific number of equal-sized groups $B$, from the PP model
defined above, with a choice of parameters given by $\theta_i=1/B$,
$\lambda_{\text{in}}=c\avg{k}N/B$ and
$\lambda_{\text{out}}=(1-c)\avg{k}N/[B(B-1)]$, with
$c=1/B+(1-1/B)\epsilon$, such that if $\epsilon=0$ we have fully random
networks, and $\epsilon=1$ we have perfectly assortative
communities. For the inference of the PP model and the DC-SBM, we sample
from the posterior distribution of Eq.~\ref{eq:posterior}, using the
algorithm above. When using the modularity function, we sample from the
target distribution
\begin{equation}
  P(\bb|\A) = \frac{\ee^{\beta Q(\bb,\A)}}{Z(\A)},
\end{equation}
where $Z(\A)=\sum_{\bb}\ee^{\beta Q(\bb,\A)}$, and $Q(\bb,\A)$ is given
by Eq.~\ref{eq:Q}. We choose $\beta = 2E\mu = 2E\ln
(\lambda_{\text{in}}/\lambda_{\text{out}})$, such that if
\begin{equation}
  \gamma = \gamma_{\text{true}} = \frac{\lambda_{\text{in}} - \lambda_{\text{out}}}{\ln \lambda_{\text{in}} - \ln \lambda_{\text{out}}},
\end{equation}
then the posterior will be proportional to the likelihood of the true
underlying model, i.e. $P(\bb|\A) \propto P(\A|\lambda_{\text{in}},
\lambda_{\text{out}}, \bm\theta^*,\bb)$. We also compare with the results obtained
with the maximum likelihood choice for $\gamma$
\begin{equation}
  \gamma = \gamma_{\text{fit}} = \frac{\lambda_{\text{in}}^* - \lambda_{\text{out}}^*}{\ln \lambda_{\text{in}}^* - \ln \lambda_{\text{out}}^*},
\end{equation}
where $\lambda_{\text{in}}^*=Be_{\text{in}}/E$ and
$\lambda_{\text{out}}^*=Be_{\text{out}}/[E(B-1)]$ (assuming
Eq.~\ref{eq:q_const} holds). Finally, we also compare with $\gamma=1$,
which corresponds to the original definition of modularity, still widely
used in practice.

\begin{figure}
  \includegraphics[width=\columnwidth]{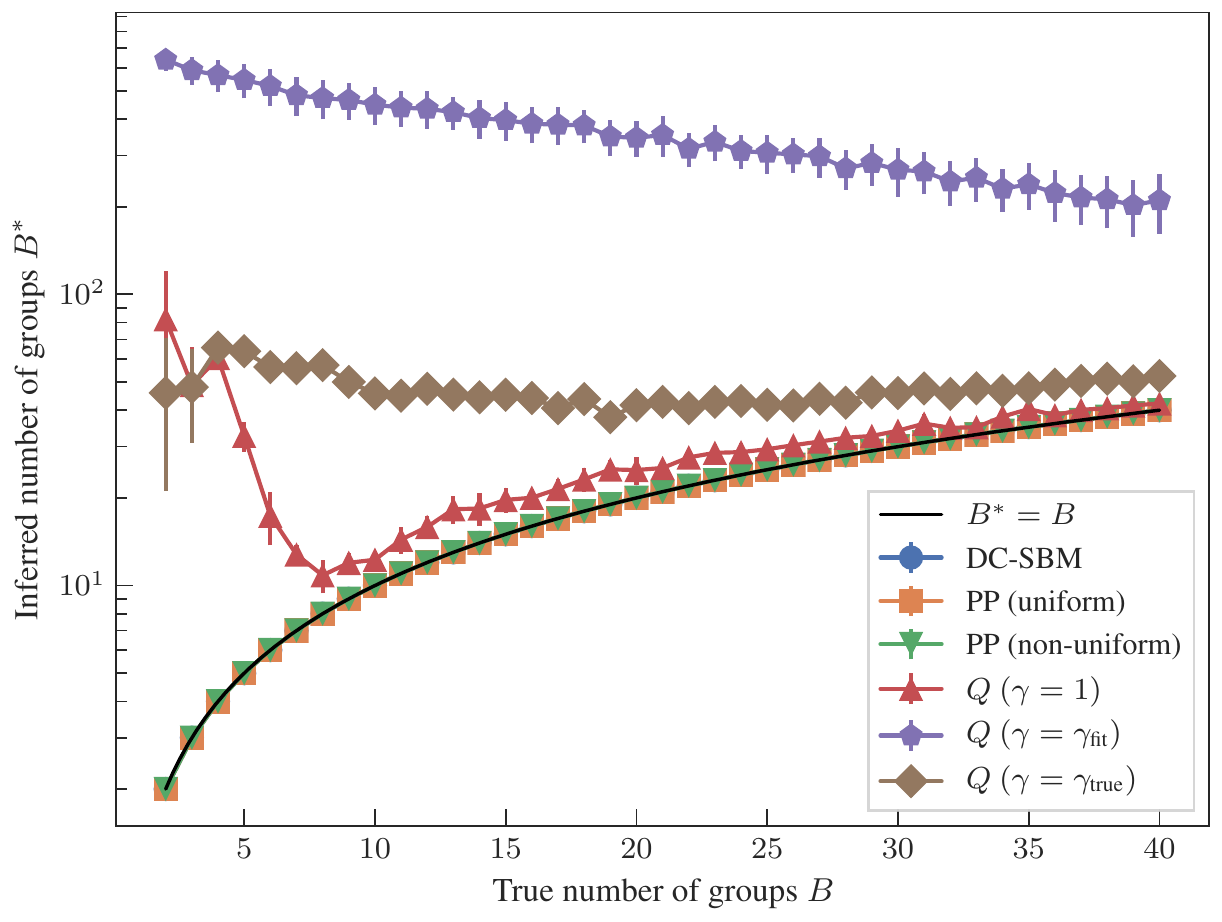} \caption{Inferred
  number of groups $B^*$, averaged from the posterior distribution, as a
  function of the true number of groups $B$, according to the procedures
  shown in the legend, for networks sampled from the PP model with
  $N=10^5$ nodes and average degree $\avg{k}=5$, and mixing
  $\epsilon=0.8$, as described in the text. The error bars show the
  standard deviation of the distribution (not its mean). The solid line
  shows the identity curve $B^*=B$. All results for the PP models
  (uniform and not) and DC-SBM are identical.\label{fig:Binf}}
\end{figure}

The results for the inferred number of groups can be seen in
Fig.~\ref{fig:Binf}. The Bayesian inference of both versions of the PP
model (uniform and non-uniform) as well as the DC-SBM yield identical
results, always returning the true number of groups. All versions of the
modularity-based approach overfit systematically, often finding a number
of groups which is orders of magnitude wrong. The bad behavior of the
case $\gamma=\gamma_{\text{true}}$ may seem surprising, since it
corresponds to the true likelihood of the model, which one could expect
to be ``Bayes optimal,'' in the sense that since it already includes the
correct model parameters other than the partition itself, then any other
approach would need to yield a strictly worse performance. However, this
would only be true if the number of groups would also be set to its true
value (rendering its inference moot), otherwise this choice of parameter
is no longer optimal. The behavior with $\gamma=\gamma_{\text{fit}}$ is
considerably worse than all others, showing how maximum likelihood is
inadequate for models with unconstrained degrees of freedom, as it
trivially overfits. Interestingly, the choice $\gamma=1$ seem to yield a
better regularization than the alternatives, although the approach still
systematically overfits, specially for a small number of planted
communities. Results like this should give us pause when employing
modularity to uncover communities in networks. Our Bayesian approach, on
the other hand, behaves robustly, without requiring us to tune any
parameter.

\subsection{Bayesian inference of the PP model has no resolution limit}\label{sec:resolution}

As was shown by Fortunato and
Barthélemy~\cite{fortunato_resolution_2007}, the method of modularity
maximization possesses an intrinsic preferred scale for the size of the
communities, which results in the so-called ``resolution limit'' that
prevents relatively small modules to be uncovered, even if they have a
very clear structure. Here we show that our Bayesian method does not
suffer from the same problem.

We begin by briefly revisiting the result of
Ref.~\cite{fortunato_resolution_2007}, and we consider the structure of
a maximally modular network, i.e. one that is constructed in order to
maximize modularity. Following Ref.~\cite{fortunato_resolution_2007} we
consider, without loss of generality, a network of $N$ nodes and $E$
edges that are divided into $B$ equal-sized groups, each with $(E-B)/B$
internal edges, connecting nodes of the same group, and in total $B$
edges connecting nodes of different communities, forming a circular ring
between communities (the ring construction simply enforces that the
network can in principle be connected, but plays no other role in the
results). With this parametrization we have $e_r=2E/B$,
$e_{\text{in}}=E-B$, $e_{\text{out}}=B$. The number of groups itself is
a free parameter, and it determines the overall modularity, which from
Eq.~\ref{eq:Q} we obtain
\begin{equation}
  Q(\A,\bb) = 1-\frac{B}{2E}-\frac{\gamma}{B}.
\end{equation}
We now seek to find the value $B=B^*$ that maximizes the above
equation. Treating $B$ as a continuous value for this purpose, and
taking the derivative and setting it to zero, $\dd Q/\dd B = 0$, we obtain
\begin{equation}
  B^* = \sqrt{2\gamma E}.
\end{equation}
This result tells us that if we construct a network in the above way but
with $B>B^*$, even if the groups themselves happen to be obvious
assortative communities, e.g. large cliques connected by single edges,
then these communities will be unintuitively merged together to achieve
a larger modularity. The above result also reveals the role of the
resolution parameter $\gamma$, which serves as the base of methods that
attempt to determine its most appropriate value to counteract the limit
in resolution~\cite{reichardt_statistical_2006,arenas_analysis_2008,
  ronhovde_multiresolution_2009,lancichinetti_limits_2011,ronhovde_local_2015}.

We now turn to the PP model, to determine if the same natural scale
emerges. We need to consider the value of $\ln P(\A,\bb)$ for the same
construction above, and determine the value of $B=B^*$ that maximizes
it. We will use the non-uniform PP model with Eqs.~\ref{eq:nupp}
and~\ref{eq:bprior}, although the final result is the same with the
uniform version. We can obtain a simpler expression for the joint
log-likelihood by assuming a large network with $N\gg 1$ and $B\gg 1$
(although we make no assumption on the value of $B$ relative to $N$ or
$E$), so that we can use Stirling's formula $\ln x! = x\ln x-x+O(\log
x)$, which yields
\begin{multline}
  \ln P(\A,\bb) = (E-N)\ln B + 2(E-B)\ln(E-B) \\
  + 2B\ln B + 2(N-B)\ln(N-B) + B \\
  - (2  E + N -B)\ln(2 E+N-B)
  + O(\ln B),
\end{multline}
up to unimportant additive constants. Taking the derivative and setting
it to zero, we obtain the equation
\begin{equation}
  \frac{\dd}{\dd B} \ln P(\A,\bb) = \frac{E-N}{B}+\ln \frac{(N+2E-B)B^2}{(E-B)^2(N-B)^2} = 0.
\end{equation}
If we now assume we have a sparse graph with $E=\avg{k}N/2$, with
$\avg{k}>2$ being the average degree independent of $N$, for $N\gg
\avg{k}$ the solution of the above equation is
\begin{equation}
  B^* = \frac{\avg{k}-2}{2} \frac{N}{\ln N}.
\end{equation}
This means that the Bayesian approach has a natural scale which prefers
group sizes $N/B^*=O(\ln N)$, which is significantly smaller than the
modularity scale $N/B^*=O(\sqrt{N})$. The scale of the Bayesian approach
arises mostly due to the requirements of statistical evidence --- we
should partition a network only if its structure cannot be explained by
a uniformly random placement of the edges. This explains also why for
$\avg{k}<2$ we obtain a value $B^*=1$, since sparser networks inherently
contain less information about the existing community
structure.\footnote{This does not mean that if $\avg{k}<2$ the modules
are completely undetectable, only that the posterior distribution has a
maximum at $B=1$. The detectability threshold, which does exist, is
determined by averaging over the posterior distribution with the true
parameters imposed~\cite{decelle_asymptotic_2011}.} As the size of the
network increases, so does the possible ways of partitioning it, and as
a consequence the required statistical evidence to support it also
increases, and hence it becomes impossible to uncover groups smaller
than $O(\ln N)$. However this threshold grows so slowly that it can
barely be compared with what exists for modularity maximization. We
emphasize that this approach virtually eliminates the resolution limit
without the introduction of a single parameter that needs to be tuned.

It is interesting to compare the value of $B^*$ for the PP model with
the same value obtained for the general SBM. As shown in
Refs.~\cite{peixoto_parsimonious_2013, peixoto_nonparametric_2017}, when
using noninformative priors, the SBM has a resolution limit
$B^*=O(\sqrt{N})$, which is similar to modularity, although it occurs
for a completely different reason, namely the model depends on a matrix
of parameters of size $O(B^2)$, which results in a penalty in the joint
log-likelihood in the order of $O(B^2\ln E)$, which becomes comparable
to the likelihood when $B\sim \sqrt{N}$ for sparse networks. This
limitation is lifted when the noninformative priors are replaced by a
sequence of nested priors and hyperpriors, resulting in the nested
SBM~\cite{peixoto_hierarchical_2014,peixoto_nonparametric_2017}, which
exhibits the natural scale $B^*=O(N/\ln N)$, similar to the PP
model. However, the PP model achieves this high resolution already with
simple noninformative priors, since it depends on a set of parameters
which has total size $O(N + B)$ in the case of the non-uniform model,
and $O(N)$ in the case of the uniform variant. This illustrates the
usefulness of simpler models, which can achieve a higher performance
than more general ones, if they happen to be a good description of the
data.

\section{Empirical networks}\label{sec:empirical}

\begin{figure}
  \includegraphics[width=\columnwidth]{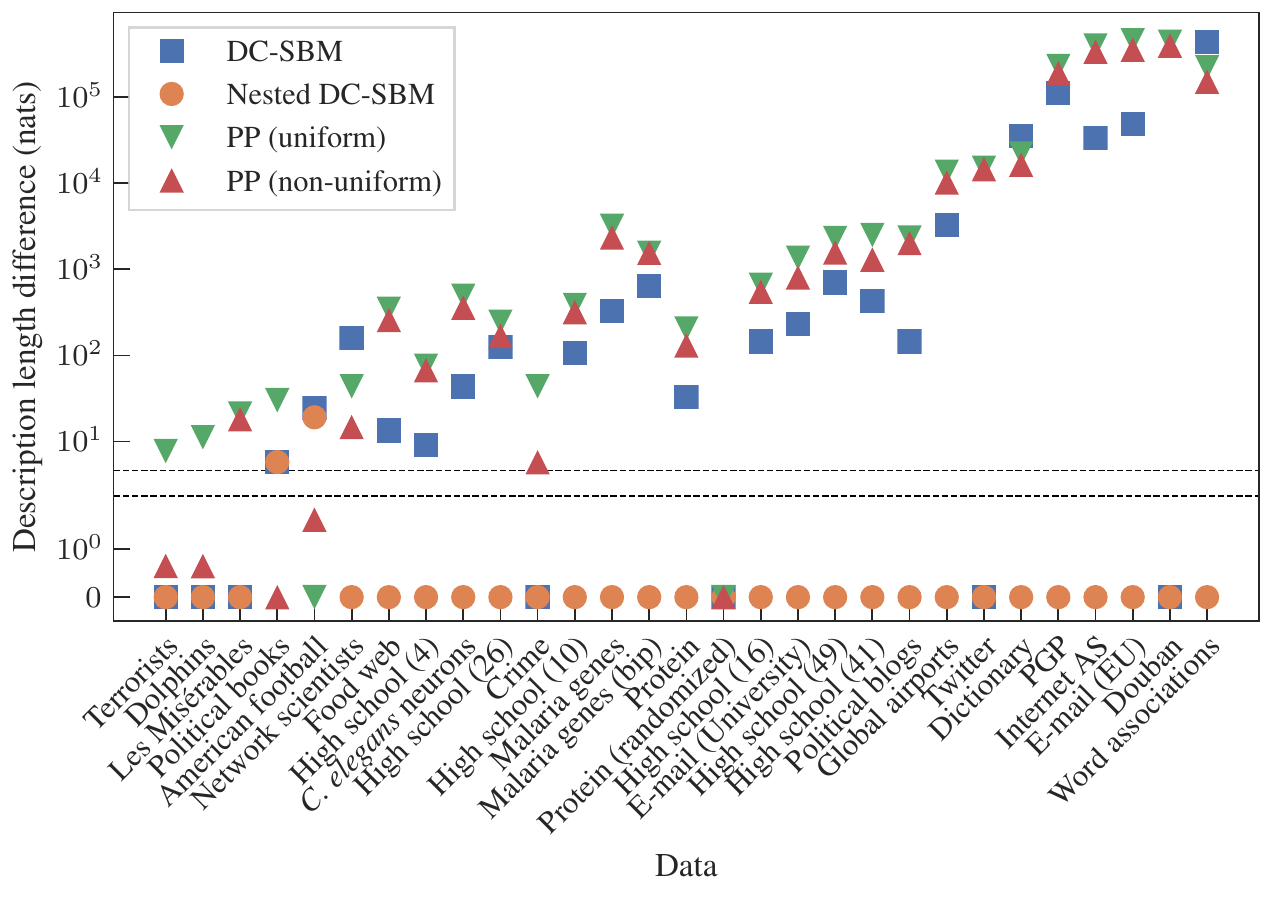}
  \caption{Difference in the description length between the best fitting
  and the remaining models, as specified in the legend, for a selection
  of networks obtained from the KONECT
  repository~\cite{kunegis_konect:_2013}. The best fitting model always
  appears in the bottom. For reference, the values $\ln 10$ and $\ln
  100$ are show as dashed lines. \label{fig:dl_diff}}
\end{figure}
\begin{figure}
  \begin{tabular}{cc}
    \multicolumn{2}{c}{American football}\\
    \includegraphics[width=.5\columnwidth]{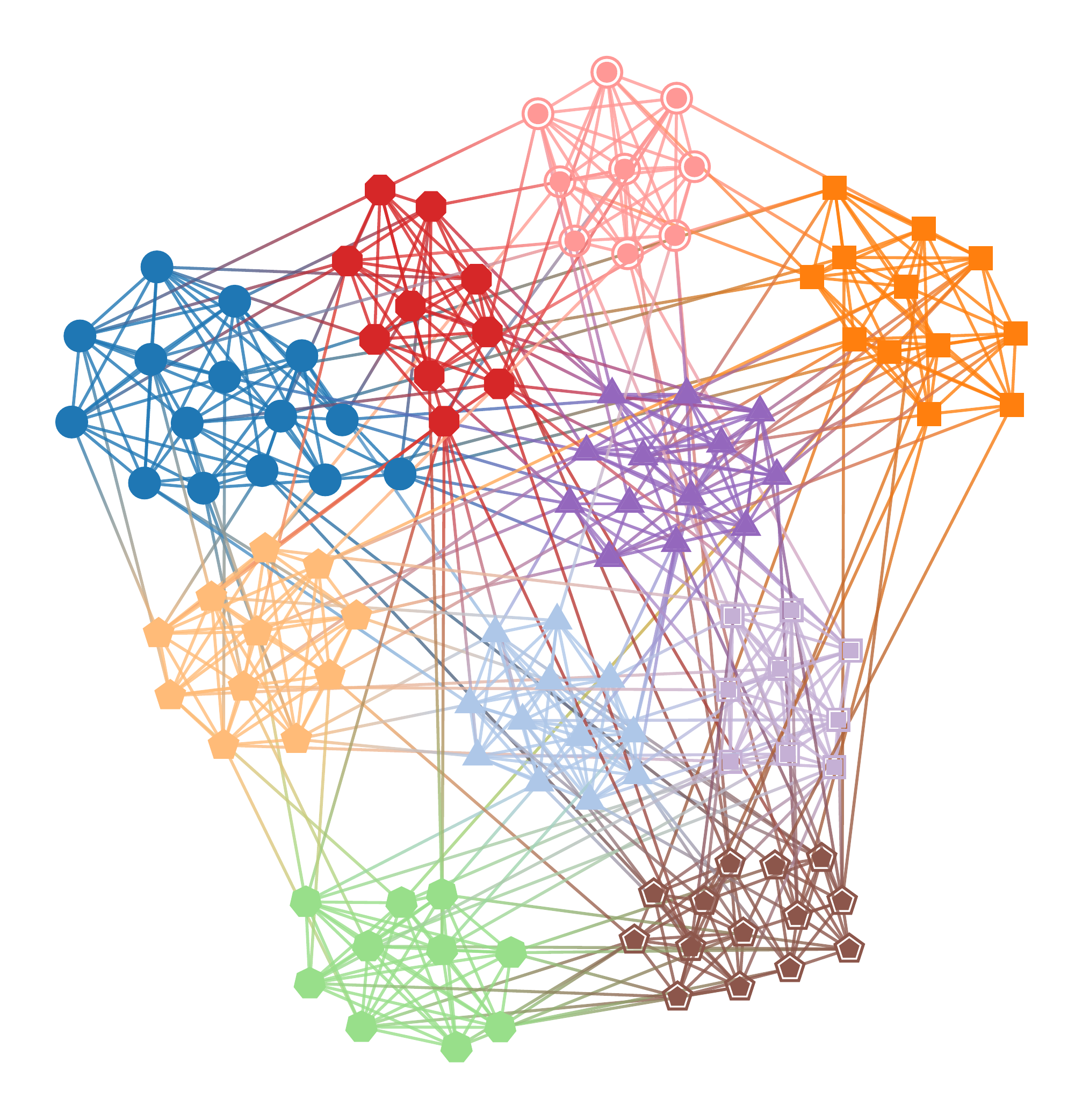}&
    \includegraphics[width=.5\columnwidth]{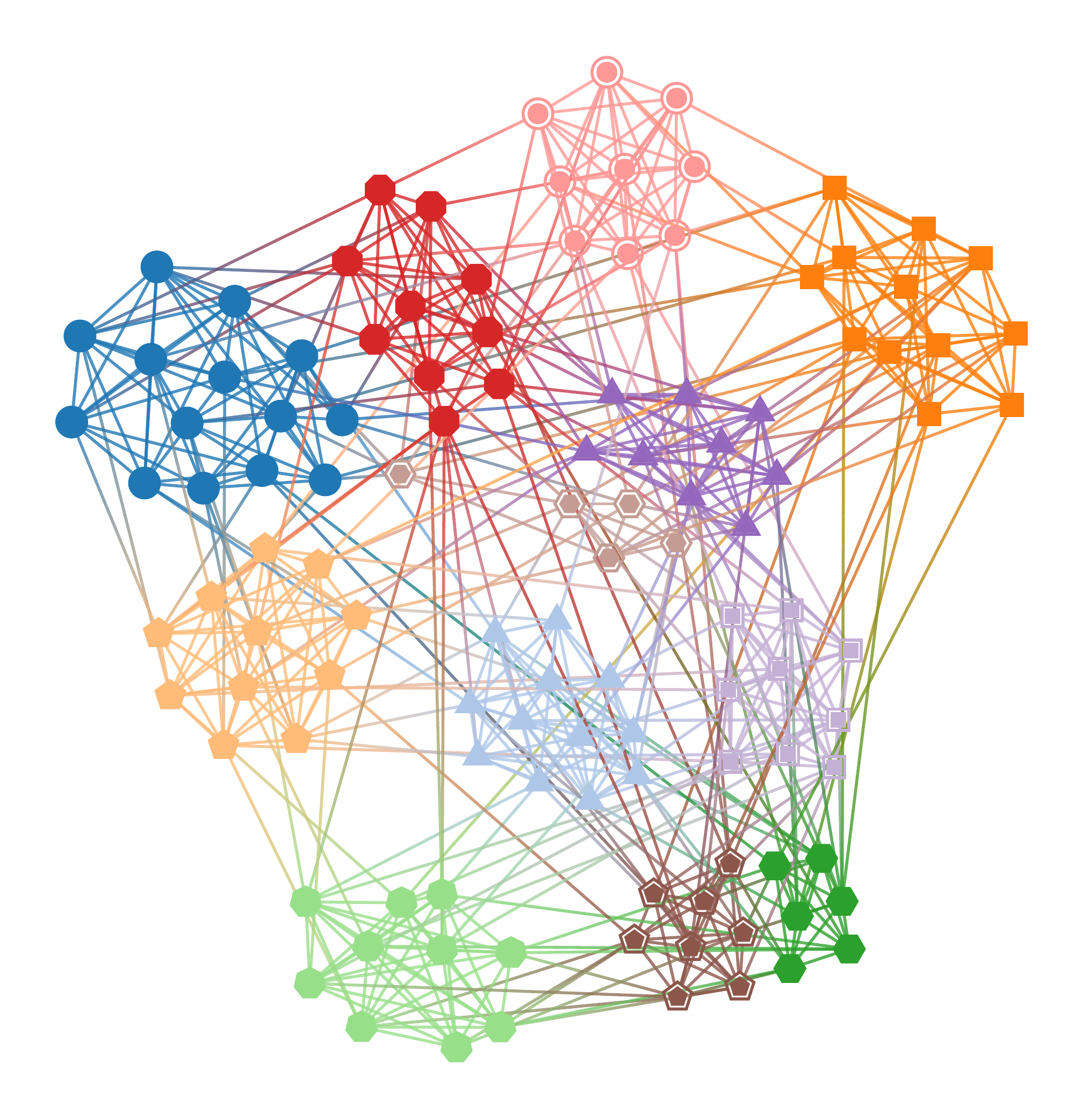}\\
    Nested DC-SBM & PP (uniform)\\
    $\Sigma=1780.58$ (nats) & $\Sigma=1761.50$ (nats)\\\rule{0pt}{1em}\\
    \multicolumn{2}{c}{Political books}\\
    \includegraphics[width=.5\columnwidth]{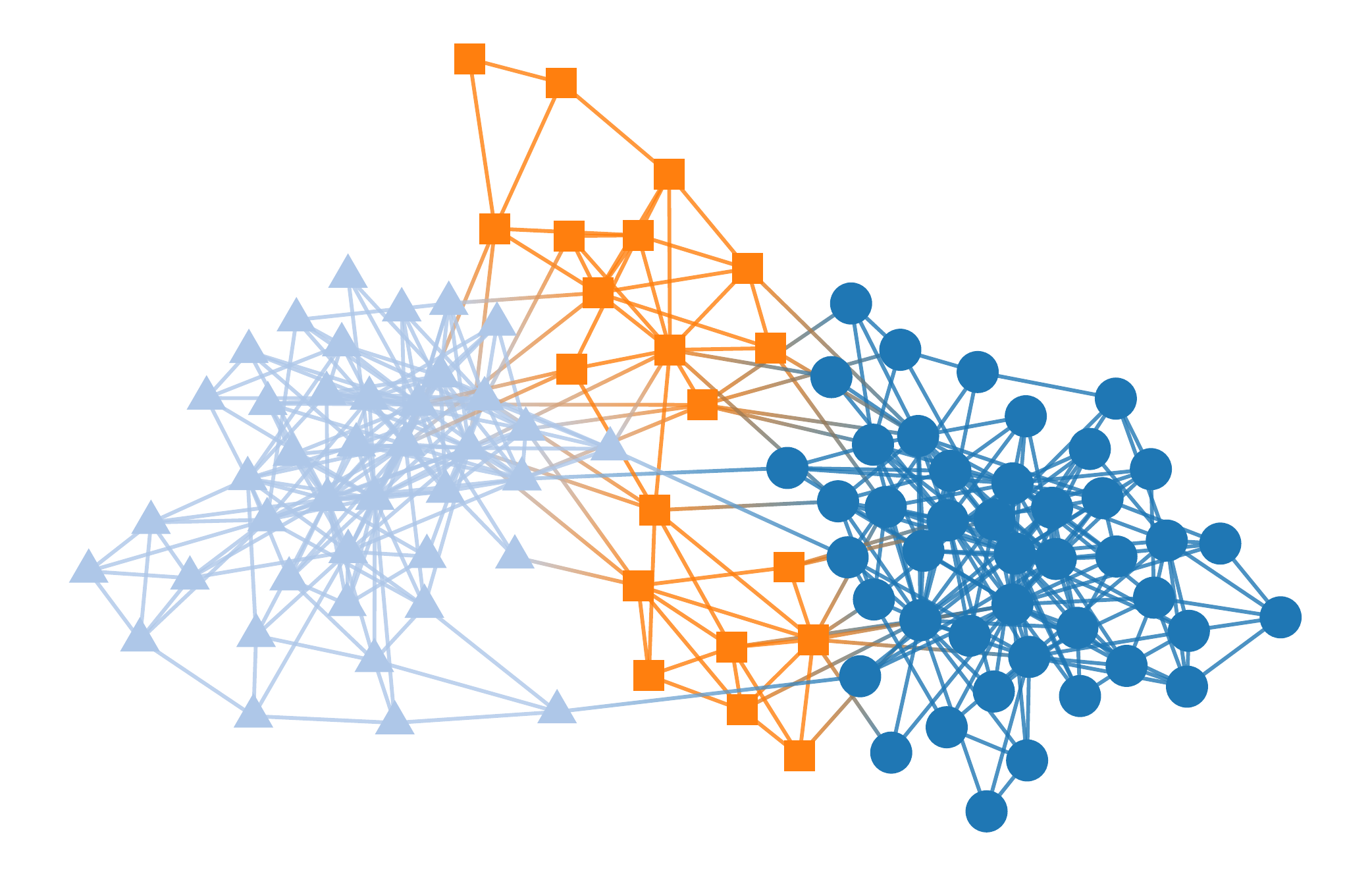}&
    \includegraphics[width=.5\columnwidth]{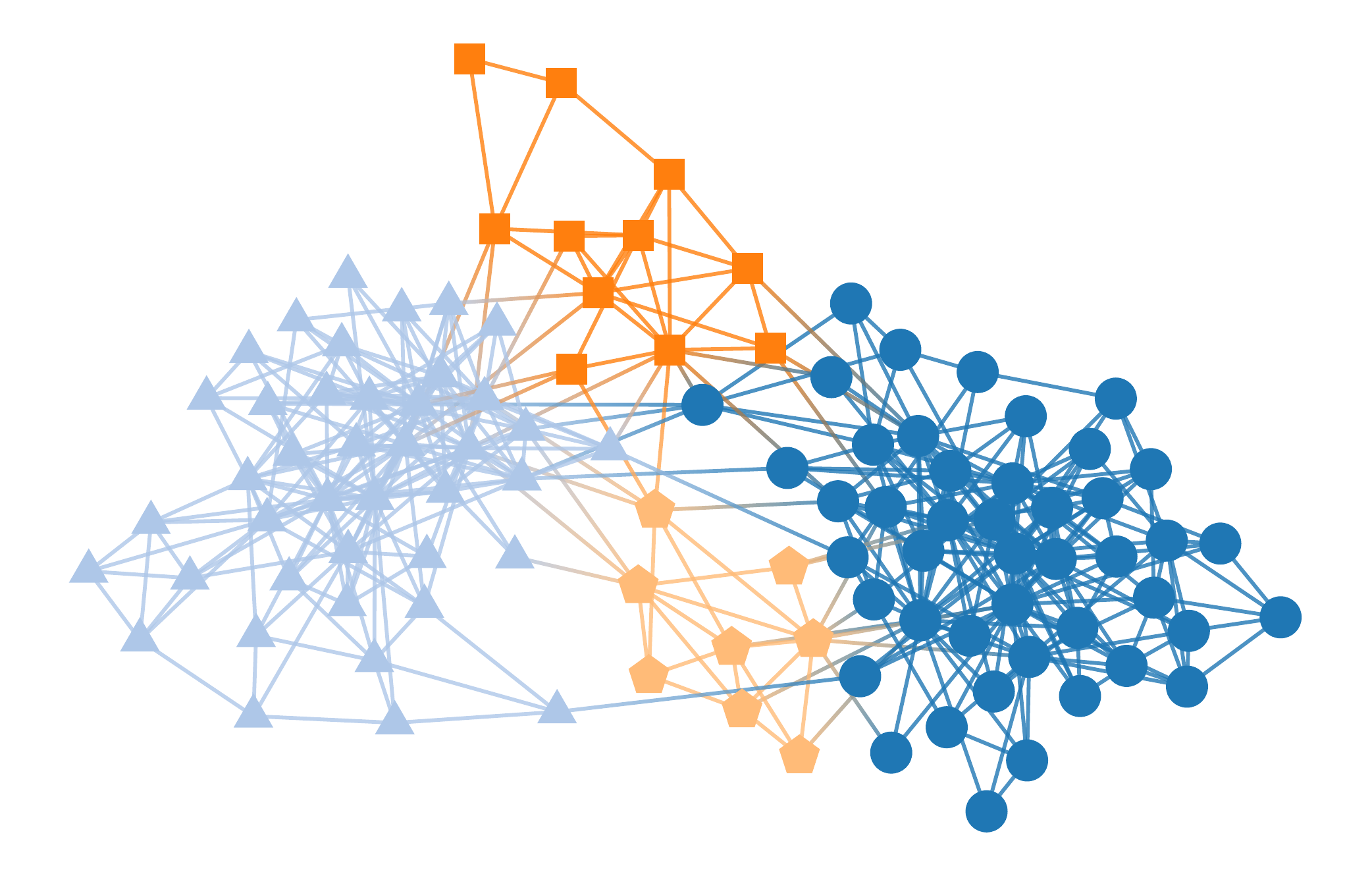}\\
    Nested DC-SBM & PP (non-uniform)\\
    $\Sigma=1343.44$ (nats) & $\Sigma=1337.69$ (nats)\\
  \end{tabular} \caption{Inferred community structure of a network of
    games between American college football
    teams~\cite{girvan_community_2002} and a network of co-purchases of
    books about American politics~\cite{krebs_political_nodate}, using
    both the PP model and the nested DC-SBM, as indicated in the legend,
    which also shows the description length of each
    fit.\label{fig:football-polbooks}}
\end{figure}

\begin{figure*}
  \begin{tabular}{cc}
    PP (non-uniform) & Nested DC-SBM \\
    \includegraphics[width=\columnwidth]{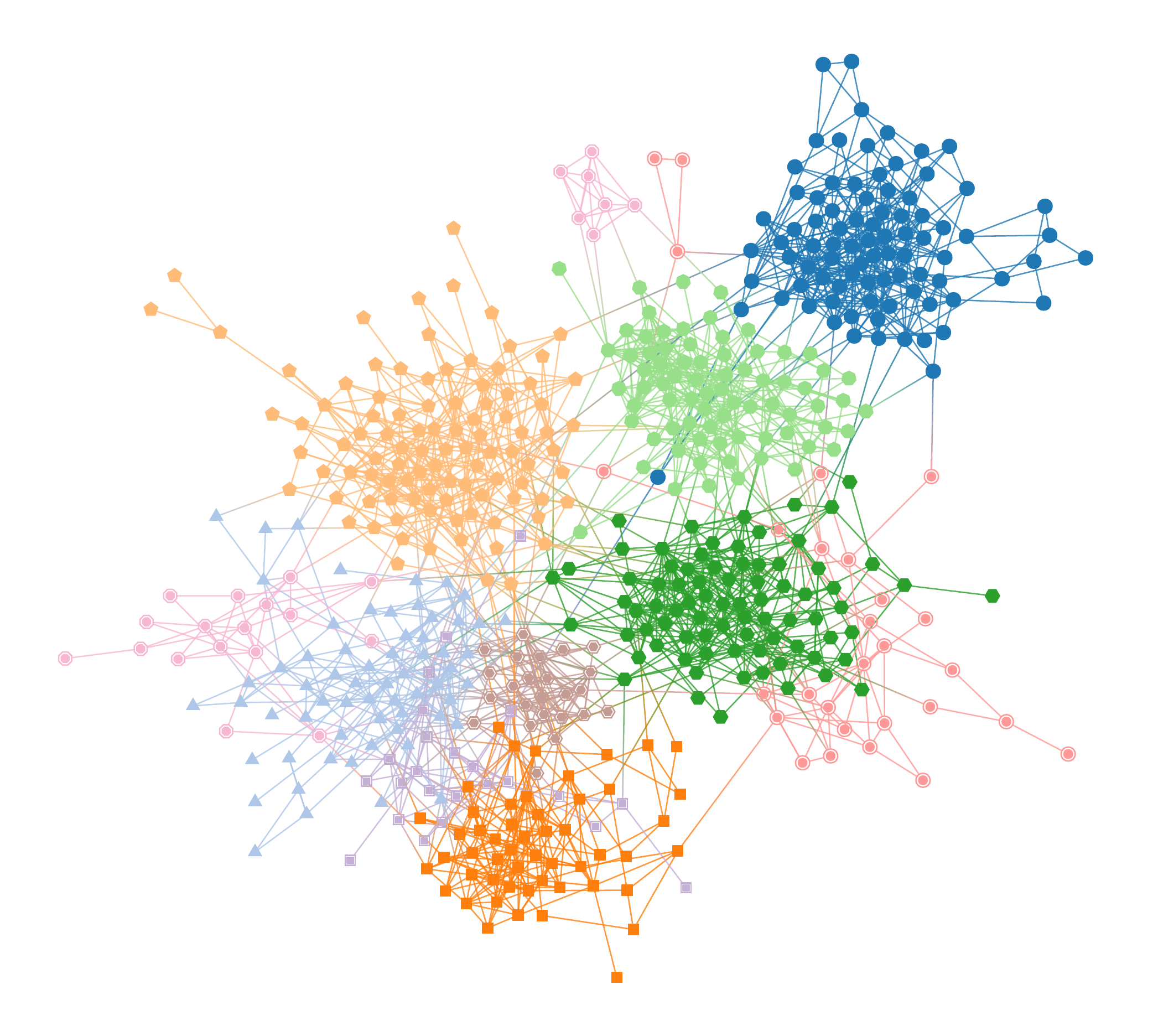} &
    \includegraphics[width=\columnwidth]{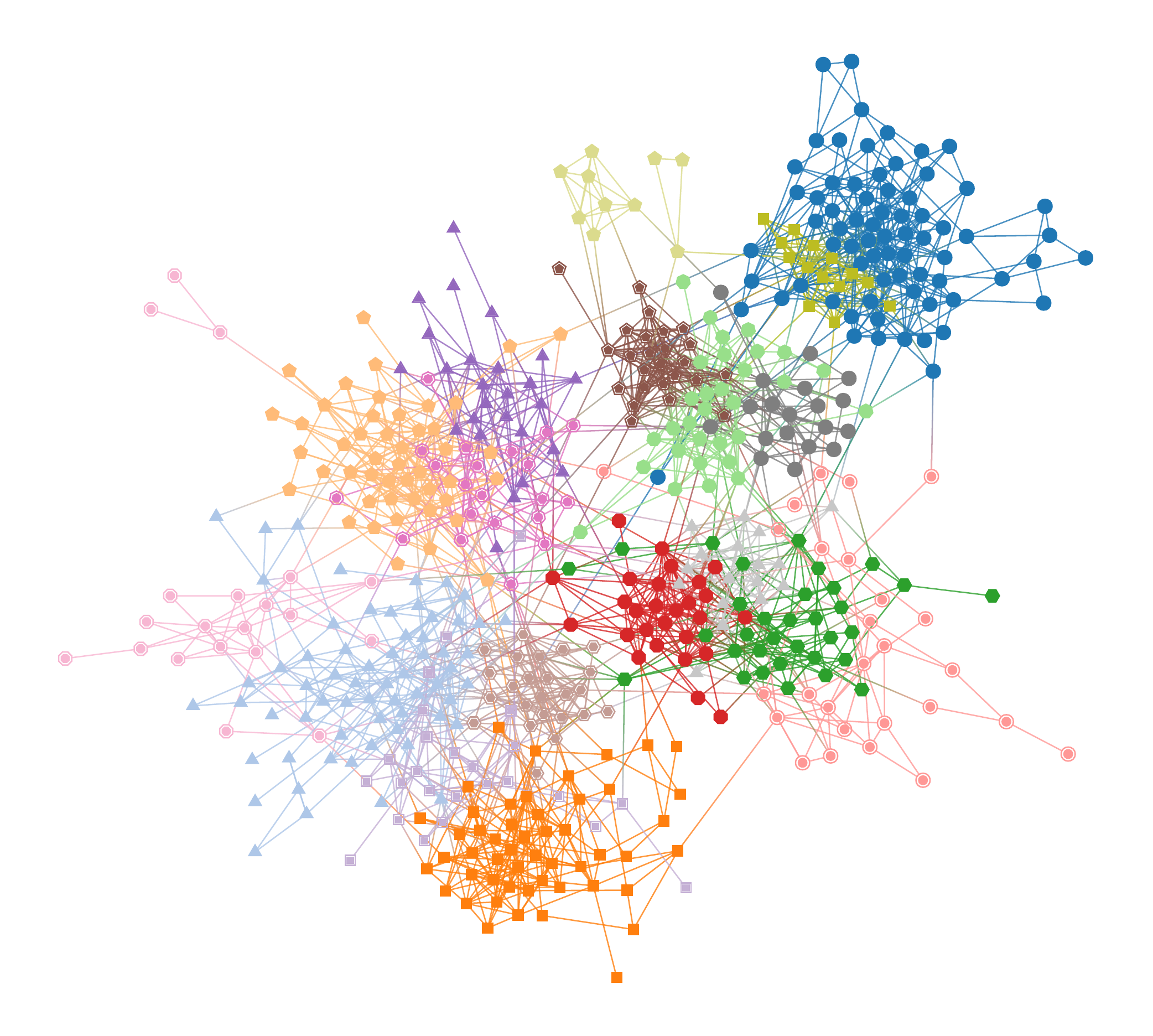} \\
    \includegraphics[width=\columnwidth]{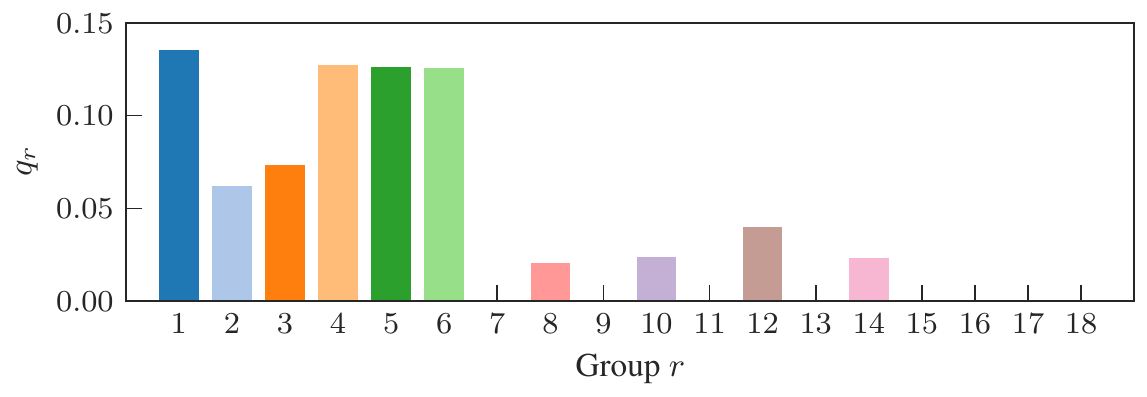} &
    \includegraphics[width=\columnwidth]{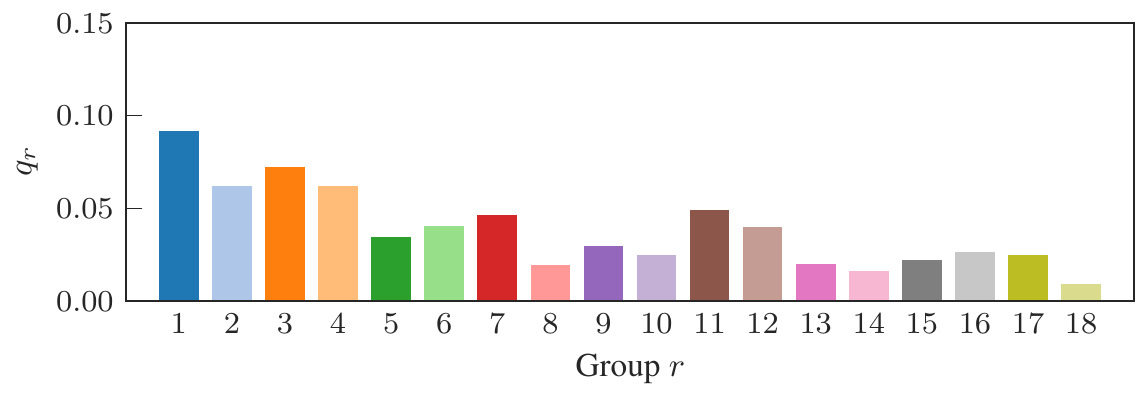} \\
    $\Sigma = 8944.09$ (nats), $Q=0.765$ &
    $\Sigma = 8775.82$ (nats), $Q=0.706$
  \end{tabular} \caption{Inferred community structure of a social
  network of high school students~\cite{harris_national_2009}, using
  both the non-uniform PP model and the nested DC-SBM. The bottom panels
  show the community-wise modularity values $q_r=(e_{rr}-e_r^2/2E)/2E$,
  such that $Q=\sum_rq_r$. A value of $q_r > 0$ indicates that group $r$
  has a predominantly assortative contribution. The bottom legend shows
  the description length obtained with both models, as well as the value
  of modularity of the partitions. Both divisions have a normalized
  maximum overlap distance~\cite{peixoto_revealing_2020} of
  $d=0.299$. The group colors are chosen to maximize the matching
  between both partitions, as described in
  Ref.~\cite{peixoto_revealing_2020}, and the same colors are used in
  the bottom panels. \label{fig:highschool}}
\end{figure*}

\begin{figure}
  \includegraphics[width=\columnwidth]{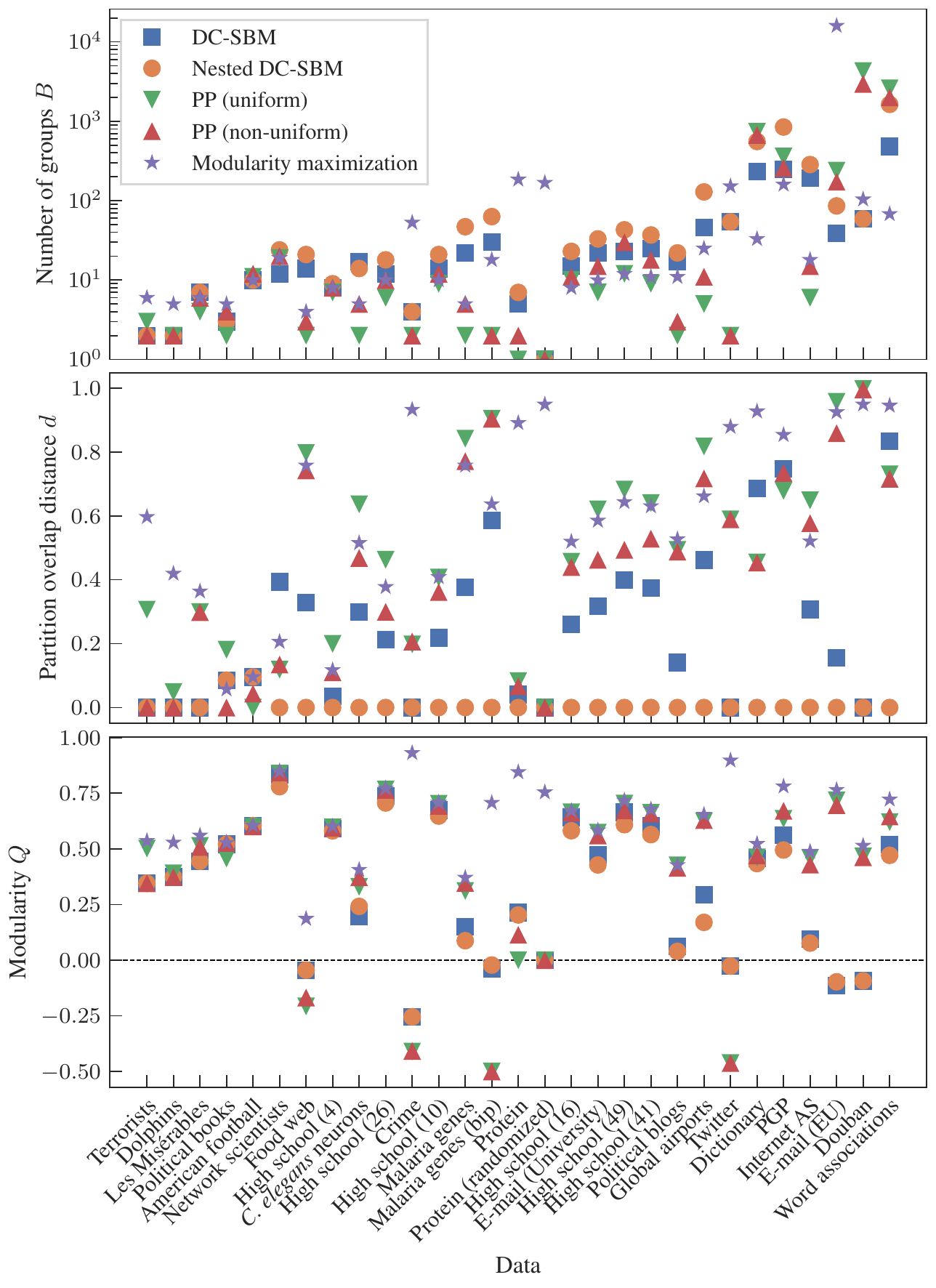} \caption{More
    details about the results of Fig.~\ref{fig:dl_diff},
    including the number of communities
    found with each method (top panel), the normalized maximum overlap
    distance~\cite{peixoto_revealing_2020} between the best fitting and the
    remaining partitions (middle panel), and the modularity of the partitions
    found (bottom panel).\label{fig:Q_dist}}
\end{figure}

The existence of assortative community structure if often assumed to be
a ubiquitous property of many kinds of networks across different
domains. However this kind of latent structure is not something that can
be readily obtained from network data, and most methods that are used to
detect it search exclusively for assortative structure, ignoring other
patterns. Therefore they cannot be used to rule out the existence of
more fundamental non-assortative mixing patterns that are qualitatively
different. A comparison between the assortative PP models that we
consider here, together with more general SBM formulations allow us to
address this comparison in a principled way, in order to understand how
pervasive assortativity really is.

Here we compare the results obtained with the inference of PP model
(both uniform and non-uniform versions) for a variety of empirical
networks, together with those obtained using a Bayesian version of the
DC-SBM~\cite{peixoto_nonparametric_2017}, using both noninformative
priors as well as nested hierarchical
priors~\cite{peixoto_hierarchical_2014}. A powerful feature of the
Bayesian inference approach is that it permits principled model
selection, in the following way. Suppose we want to compare the
community structure $\bb_1$ found with model $\mathcal{M}_1$ with
structure $\bb_2$ found with model $\mathcal{M}_2$, both for the same
network $\A$. We can do this by comparing their posterior probability
ratio
\begin{align}
  \Lambda
  &= \frac{P(\bb_1,\mathcal{M}_1|\A)}{P(\bb_2,\mathcal{M}_2|\A)}\\
  &= \frac{P(\A,\bb_1|\mathcal{M}_1)P(\mathcal{M}_1)}{P(\A,\bb_2|\mathcal{M}_2)P(\mathcal{M}_2)}.
\end{align}
Therefore, if we have no prior preference towards any model,
i.e. $P(\mathcal{M}_1)=P(\mathcal{M}_2)$, then this ratio will be given
by the difference in the description length obtained with both models,
\begin{equation}
  \Lambda = \exp(\Sigma_2 - \Sigma_1),
\end{equation}
where $\Sigma_1=-\ln P(\A,\bb_1|\mathcal{M}_1)$ and $\Sigma_2=-\ln
P(\A,\bb_2|\mathcal{M}_2)$. Hence, the most likely model is the one that
offers the best compression for the data, and the difference in the
compression itself yields the statistical significance of the preference
towards the best model.

We performed the inference of the four models on a selection of 29
networks, representing different scientific domains, obtained from the
KONECT repository~\cite{kunegis_konect:_2013}. In Fig.~\ref{fig:dl_diff}
we show the difference in description length obtained from the best
fitting to the other models. Perhaps unsurprisingly, we find that the
general DC-SBM provides a better fit for most networks, indicating that
the strictly assortative structure of the PP model is insufficient to
account for the observed networks. However, the PP model is selected as
the best fitting model in a minority of the cases, and it is instructive
to inspect those more closely. In Fig.~\ref{fig:football-polbooks} we
show the communities uncovered using the PP model and the nested DC-SBM,
for a network of games between American college football
teams~\cite{girvan_community_2002} and a network of co-purchases of
books about American politics~\cite{krebs_political_nodate}. In both
cases, PP and the nested DC-SBM find very similar partitions, but the PP
model finds a slightly larger number of communities. The model selection
criterion outlined above selects the PP model as the more plausible
alternative due to the strong assortativity observed. The result found for
the football network is particularly interesting, since it is a rare
case where the \emph{uniform} PP model is the one that gets
selected. This is because the number of edges inside each community is
indeed very similar for all of them, and the connections between the
communities seem fairly random, exactly how the PP model
prescribes. This highlights the robust character of our approach, which
will not favor a more complicated model when it is unnecessary, and
gives us confidence that when the PP model is not selected, it is indeed
because it does not fully account for the actual structure observed in
the network.

For other networks such as the associations between
terrorists~\cite{krebs_uncloaking_2002} and the social network between
dolphins~\cite{lusseau_bottlenose_2003}, even though the DC-SBM is
strictly preferred, the difference between the nonuniform PP model is
negligible, and therefore there is no sufficient evidence in the data to
reliable distinguish between both models. For all other data, however,
we find substantial evidence in favor of the more general DC-SBM. What
is particularly interesting is that the DC-SBM is often preferred even
when the uncovered structures are in fact very assortative. We give an
example of this in Fig.~\ref{fig:highschool}, which shows the
communities found with the non-uniform PP model and the nested DC-SBM
for a social network of high school
students~\cite{harris_national_2009}. Even though all communities found
have a larger probability of forming internal than external connections,
the ones found by the DC-SBM yield a larger plausibility. If we inspect
it more closely, we see that the divisions found by the DC-SBM amount
largely to subdivisions of the ones found by the PP model. This can be
explained by the DC-SBM using the preference of connections
\emph{between} the different communities as additional evidence for
their existence, instead of merely their assortativity strength. This
illustrates how a more general model like the DC-SBM can be more useful
even when assortativity is a dominant but not unique pattern. (We note
that the result found by the PP model has a slightly larger modularity,
but this is not a very significant fact, given that modularity is in
general largely decoupled form statistical significance.)

In Fig.~\ref{fig:Q_dist} we show more details about the inferences
obtained for all the networks, including the number of communities
found, the normalized maximum overlap
distance~\cite{peixoto_revealing_2020} between the best fitting and the
remaining partitions, and the modularity of the partitions
found. Overall, we observe a fair amount of variability in the
comparisons between models for the different networks. Very often, the
PP models yield a more conservative view of the networks, uncovering a
smaller number of groups when compared to the DC-SBM, but there are also
cases where the opposite is true. We also observe that although there
are many cases where both the DC-SBM and PP models yield partitions with
similar modularity, the overlap distance between partitions is very
high, indicating that these networks admit a variety of divisions that
have a similar overall level of assortativity (a good example of this is
the high school network we considered in
Fig.~\ref{fig:highschool}). Therefore, despite similar values of
modularity found with the DC-SBM, the more general model rarely yields
partitions that are very similar to the ones returned by any of the PP
variants.

The values of modularity obtained with the best fitting model (which is
most often the nested DC-SBM) are in some situations similar to what is
found with the PP models, like for the high school social
networks. However, for networks like the \texttt{douban.com} online
social recommendation network~\cite{reza_social_2009}, political
blogs~\cite{adamic_political_2005}, Internet at the autonomous system
level~\cite{peixoto_hierarchical_2014}, and others, the modularity
obtained with the best-fitting DC-SBM is significantly smaller than with
the PP models, indicating that assortativity is not the most fundamental
pattern in these networks, and using a community detection method that
searches exclusively for these patterns gives us a significantly biased
view.

\begin{figure}
  \begin{tabular}{cc}
    \multicolumn{2}{c}{Original network}\\
    \includegraphics[width=.5\columnwidth]{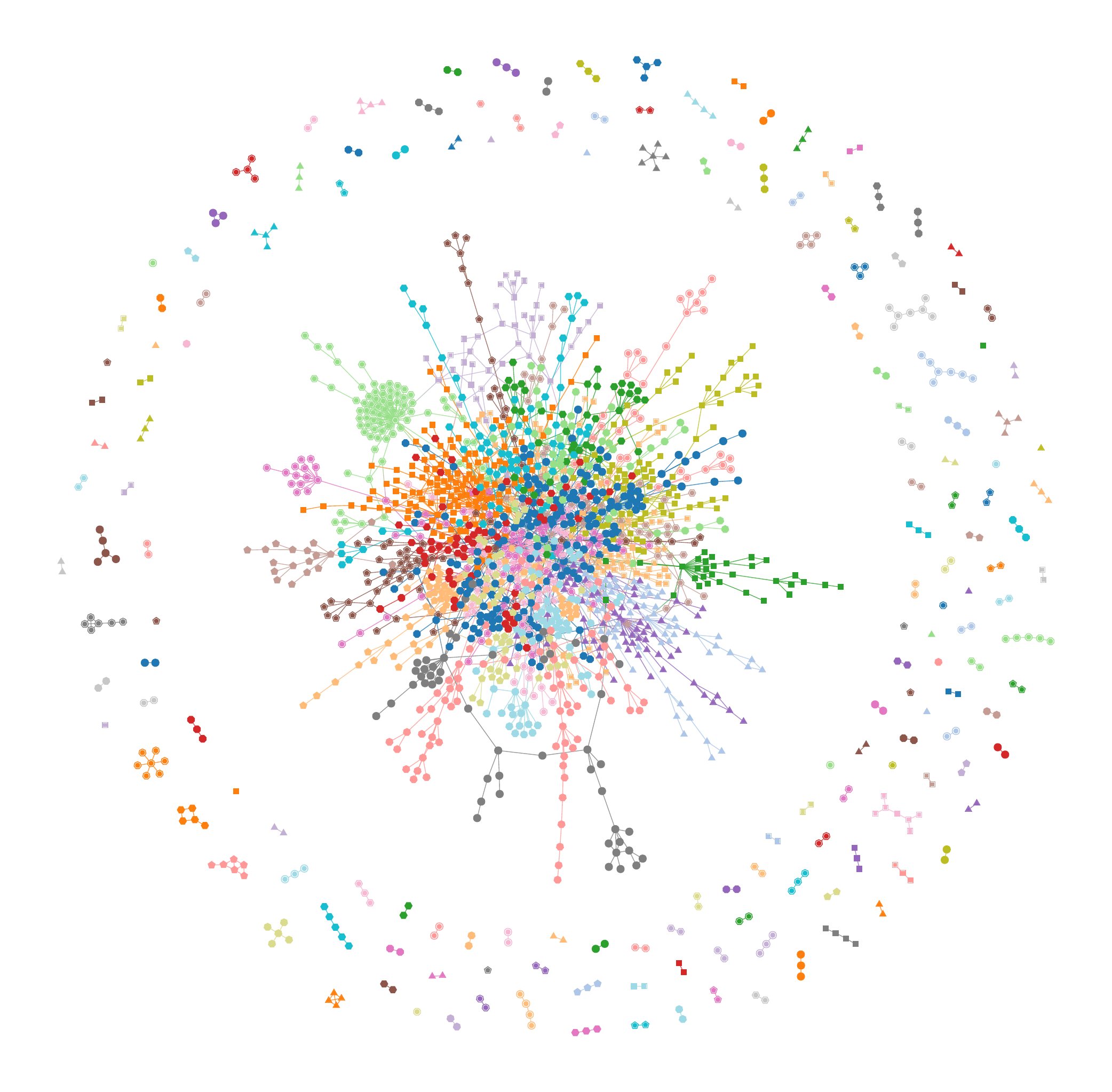} &
    \includegraphics[width=.5\columnwidth]{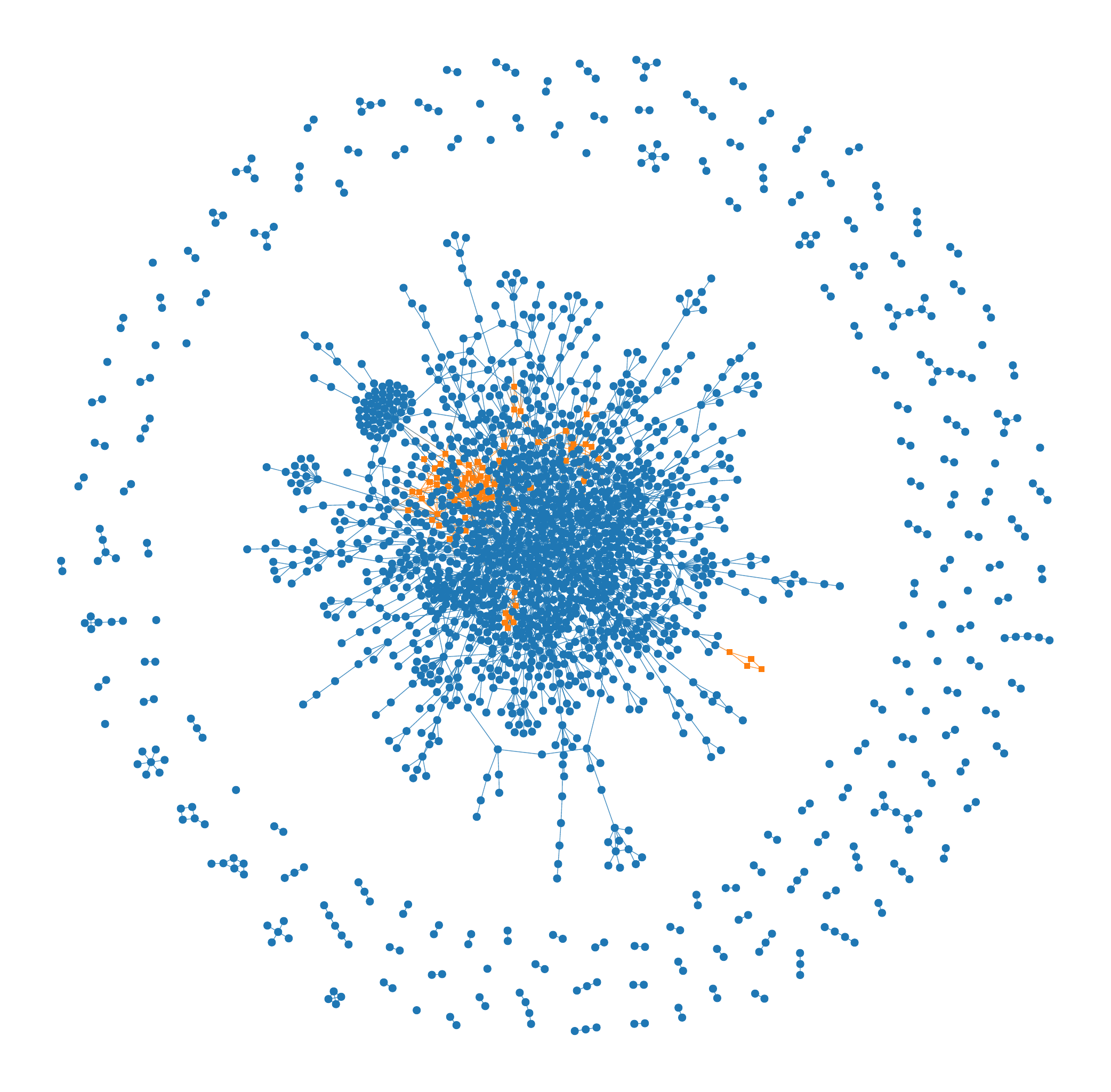} \\
    Modularity maximization & PP (non-uniform)\\
    $B=185$, $Q=0.84$ & $B=2$, $Q=0.11$\\\rule{0pt}{1em}\\
    \multicolumn{2}{c}{Randomized network}\\
    \includegraphics[width=.5\columnwidth]{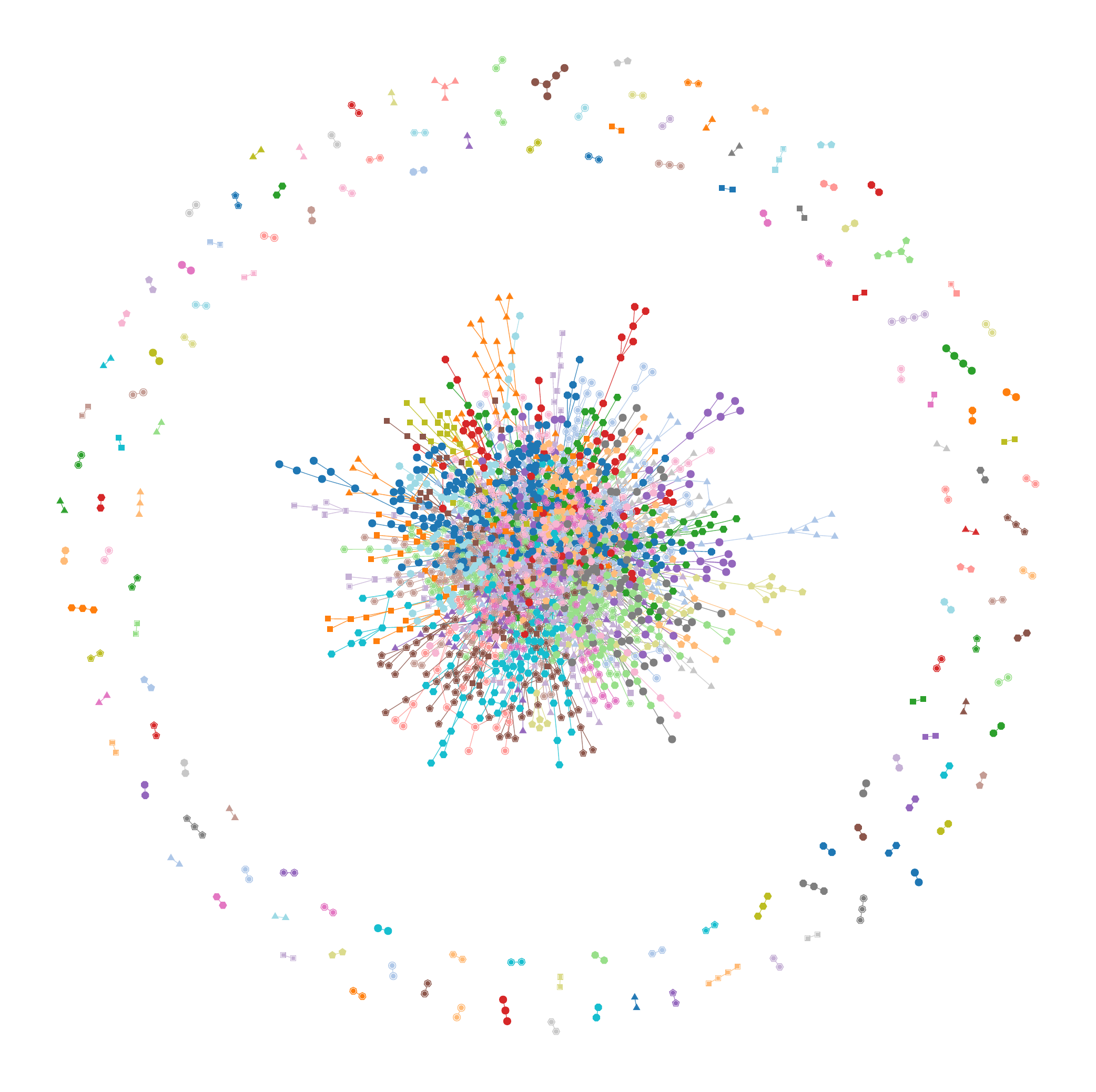} &
    \includegraphics[width=.5\columnwidth]{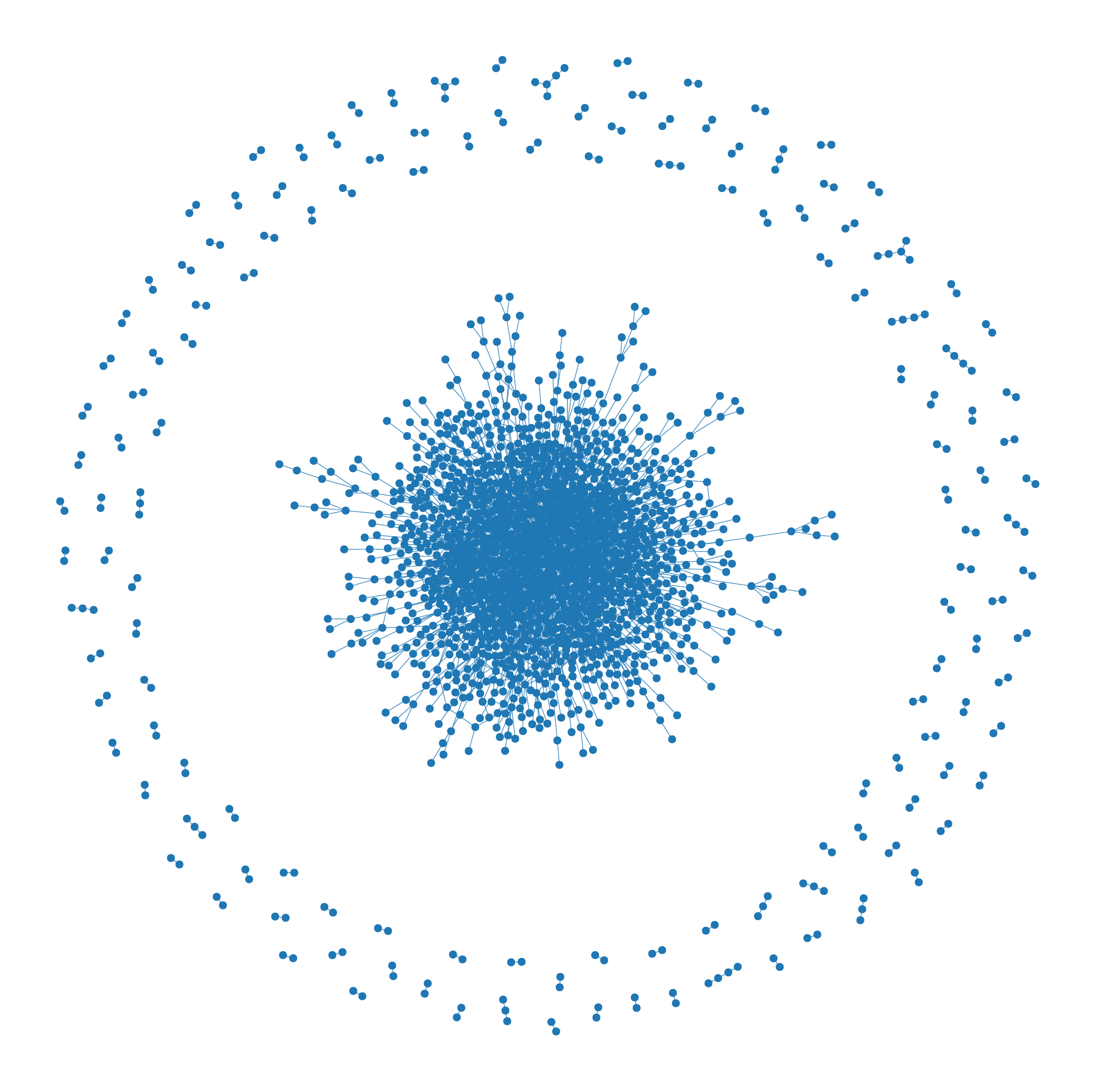} \\
    Modularity maximization & PP (non-uniform)\\
    $B=168$, $Q=0.75$ & $B=1$, $Q=0$
  \end{tabular}

  \caption{Community structures found in a network of protein-protein
    interactions~\cite{jeong_lethality_2001}, using modularity
    maximization and Bayesian inference of the PP model, as indicated in
    the legends. The top panel shows the results for the original
    network, and in the bottom panel the results obtained for a
    randomized version with the same degree sequence. The legends show
    the number of groups and the value of modularity of the
    corresponding partitions.\label{fig:protein}}
\end{figure}

In Fig.~\ref{fig:Q_dist} we also include results obtained with the
method of modularity maximization. As it must be the case, this approach
yields the division of the network with the highest values of modularity
among the alternative ones considered. When we compare the results
obtained with more statistically grounded approaches, we observe a
rather erratic behavior. For some networks, such as word
associations~\cite{fellbaum_wordnet_1998}, modularity maximization
yields a seemingly more conservative result with fewer groups, which
could be an underfit potentially due to the resolution
limit~\cite{fortunato_resolution_2007}. In other instances, like the
E-mail network of an undisclosed European
institution~\cite{leskovec_graph_2007}, protein-protein
interactions~\cite{jeong_lethality_2001} and bipartite person-crime
associations~\cite{decker_st_1991}, modularity maximization finds a
number of communities that is multiple orders of magnitude larger than
what is obtained with the inference methods, strongly indicating a
massive overfit of these datasets. We illustrate this point in more
detail by focusing on the protein-protein interaction network in
Fig.~\ref{fig:protein}. There we see that while modularity maximization
finds over a hundred communities, the inference of the PP model finds
only two, with one of them being relatively small. If we now consider a
fully randomized version of the network, shown in Fig.~\ref{fig:protein}
as well, we see that modularity maximization still finds a very similar
number of communities in it, with a high value of modularity, while the
inference of the PP model finds, correctly, only a single group. This
example clearly shows that while the structure of the original network
is probably not completely random, most of it (including its
disconnectedness) can be explained by its degree sequence alone, with no
convincing evidence of community structure, and that the results
obtained with modularity maximization are mostly spurious.

Our findings corroborate a recent analysis based on link prediction,
carried over a large corpus of empirical networks, that showed that
modularity maximization tends to systematically
overfit~\cite{ghasemian_evaluating_2019}. Together with our results,
this serves to illustrate that the tenuous connection with maximum
likelihood of the PP model should not encourage practitioners to employ
modularity maximization in the analysis of real networks, if they expect
to be guided by statistical significance, or have any inherent guarantee
against overfitting or underfitting.

\section{Conclusion}\label{sec:conclusion}

We have described how to perform a nonparametric Bayesian inference of
the planted partition generative model, resulting in a principled
community detection algorithm tailored for assortative structures. Our
method separates structure from randomness, and does not find spurious
communities in fully random networks. We also showed that it does not
suffer from the resolution limit present in modularity-based methods,
and is capable of uncovering arbitrarily small communities, provided
those are statistically significant, and without the tuning of any
parameter.  Our approach is based on the sampling or a maximization of a
posterior probability function that is not much more complicated to
implement than popular heuristics like modularity, and hence can be used
as a drop-in replacement for it in a variety of algorithms,
intrinsically providing better regularization for them.

We showed how our inference approach is amenable to statistical model
selection, and we have compared our model variations, together with more
general stochastic block models, on a variety of empirical networks. We
discussed how this comparison allows us to determine the true
assortativity of community structures, by removing a systemic bias that
exists when only constrained methods are employed. We have shown that in
many cases the assortativity of real networks is exaggerated when viewed
through the lenses of community detection methods that search
exclusively for assortative patterns, and how model selection can reveal
more fundamental mixing patterns.

A recent investigation of a variety of community detection methods on a
large corpus of empirical networks found that most of them tend to yield
a number of communities that seems compatible with an overall
$O(\sqrt{N})$ scaling~\cite{ghasemian_evaluating_2019}, indicating that
this might be a limitation that is present in a larger set of community
detection algorithms. That analysis did not include statistical
inference methods that are known not to have this particular limitation,
like the nested
SBM~\cite{peixoto_hierarchical_2014,peixoto_nonparametric_2017} and our
Bayesian PP model, as we have demonstrated. Incorporating these methods
into such large-scale comparisons would allow us to better understand
what are the true fundamental limitations of the community detection
task.

The inference approach is infinitely extensible, as it admits any
conceivable generative model, and it provides a general platform for a
meaningful comparison between them. It is easy to envision a more
general comparison across network models that are tailored towards other
kinds of specific mixing patterns, such as
bipartiteness~\cite{yen_community_2020} and
core-peripheries~\cite{zhang_identification_2015,gallagher_clarified_2020},
as well as different classes of models such as those based on latent
spaces~\cite{boguna_sustaining_2010,newman_generalized_2015-1}. A
systematic comparison under such a framework would shed important light
on the inherent trade-offs between more general and specific models, and
how they relate to the various empirical domains.

\bibliography{bib}

\begin{thebibliography}{58}%
\makeatletter
\providecommand \@ifxundefined [1]{%
 \@ifx{#1\undefined}
}%
\providecommand \@ifnum [1]{%
 \ifnum #1\expandafter \@firstoftwo
 \else \expandafter \@secondoftwo
 \fi
}%
\providecommand \@ifx [1]{%
 \ifx #1\expandafter \@firstoftwo
 \else \expandafter \@secondoftwo
 \fi
}%
\providecommand \natexlab [1]{#1}%
\providecommand \enquote  [1]{``#1''}%
\providecommand \bibnamefont  [1]{#1}%
\providecommand \bibfnamefont [1]{#1}%
\providecommand \citenamefont [1]{#1}%
\providecommand \href@noop [0]{\@secondoftwo}%
\providecommand \href [0]{\begingroup \@sanitize@url \@href}%
\providecommand \@href[1]{\@@startlink{#1}\@@href}%
\providecommand \@@href[1]{\endgroup#1\@@endlink}%
\providecommand \@sanitize@url [0]{\catcode `\\12\catcode `\$12\catcode
  `\&12\catcode `\#12\catcode `\^12\catcode `\_12\catcode `\%12\relax}%
\providecommand \@@startlink[1]{}%
\providecommand \@@endlink[0]{}%
\providecommand \url  [0]{\begingroup\@sanitize@url \@url }%
\providecommand \@url [1]{\endgroup\@href {#1}{\urlprefix }}%
\providecommand \urlprefix  [0]{URL }%
\providecommand \Eprint [0]{\href }%
\providecommand \doibase [0]{http://dx.doi.org/}%
\providecommand \selectlanguage [0]{\@gobble}%
\providecommand \bibinfo  [0]{\@secondoftwo}%
\providecommand \bibfield  [0]{\@secondoftwo}%
\providecommand \translation [1]{[#1]}%
\providecommand \BibitemOpen [0]{}%
\providecommand \bibitemStop [0]{}%
\providecommand \bibitemNoStop [0]{.\EOS\space}%
\providecommand \EOS [0]{\spacefactor3000\relax}%
\providecommand \BibitemShut  [1]{\csname bibitem#1\endcsname}%
\let\auto@bib@innerbib\@empty
%</preamble>
\bibitem [{\citenamefont {Fortunato}(2010)}]{fortunato_community_2010}%
  \BibitemOpen
  \bibfield  {author} {\bibinfo {author} {\bibfnamefont {Santo}\ \bibnamefont
  {Fortunato}},\ }\bibfield  {title} {\enquote {\bibinfo {title} {Community
  detection in graphs},}\ }\href {\doibase 16/j.physrep.2009.11.002} {\bibfield
   {journal} {\bibinfo  {journal} {Physics Reports}\ }\textbf {\bibinfo
  {volume} {486}},\ \bibinfo {pages} {75--174} (\bibinfo {year}
  {2010})}\BibitemShut {NoStop}%
\bibitem [{\citenamefont {Fortunato}\ and\ \citenamefont
  {Hric}(2016)}]{fortunato_community_2016}%
  \BibitemOpen
  \bibfield  {author} {\bibinfo {author} {\bibfnamefont {Santo}\ \bibnamefont
  {Fortunato}}\ and\ \bibinfo {author} {\bibfnamefont {Darko}\ \bibnamefont
  {Hric}},\ }\bibfield  {title} {\enquote {\bibinfo {title} {Community
  detection in networks: {A} user guide},}\ }\href {\doibase
  10.1016/j.physrep.2016.09.002} {\bibfield  {journal} {\bibinfo  {journal}
  {Physics Reports}\ } (\bibinfo {year} {2016}),\
  10.1016/j.physrep.2016.09.002}\BibitemShut {NoStop}%
\bibitem [{\citenamefont {Wasserman}\ and\ \citenamefont
  {Anderson}(1987)}]{wasserman_stochastic_1987}%
  \BibitemOpen
  \bibfield  {author} {\bibinfo {author} {\bibfnamefont {Stanley}\ \bibnamefont
  {Wasserman}}\ and\ \bibinfo {author} {\bibfnamefont {Carolyn}\ \bibnamefont
  {Anderson}},\ }\bibfield  {title} {\enquote {\bibinfo {title} {Stochastic a
  posteriori blockmodels: {Construction} and assessment},}\ }\href {\doibase
  16/0378-8733(87)90015-3} {\bibfield  {journal} {\bibinfo  {journal} {Social
  Networks}\ }\textbf {\bibinfo {volume} {9}},\ \bibinfo {pages} {1--36}
  (\bibinfo {year} {1987})}\BibitemShut {NoStop}%
\bibitem [{\citenamefont {Reichardt}\ and\ \citenamefont
  {White}(2007)}]{reichardt_role_2007}%
  \BibitemOpen
  \bibfield  {author} {\bibinfo {author} {\bibfnamefont {J.}~\bibnamefont
  {Reichardt}}\ and\ \bibinfo {author} {\bibfnamefont {D.~R.}\ \bibnamefont
  {White}},\ }\bibfield  {title} {\enquote {\bibinfo {title} {Role models for
  complex networks},}\ }\href {\doibase 10.1140/epjb/e2007-00340-y} {\bibfield
  {journal} {\bibinfo  {journal} {The European Physical Journal B}\ }\textbf
  {\bibinfo {volume} {60}},\ \bibinfo {pages} {217--224} (\bibinfo {year}
  {2007})}\BibitemShut {NoStop}%
\bibitem [{\citenamefont {Karrer}\ and\ \citenamefont
  {Newman}(2011)}]{karrer_stochastic_2011}%
  \BibitemOpen
  \bibfield  {author} {\bibinfo {author} {\bibfnamefont {Brian}\ \bibnamefont
  {Karrer}}\ and\ \bibinfo {author} {\bibfnamefont {M.~E.~J.}\ \bibnamefont
  {Newman}},\ }\bibfield  {title} {\enquote {\bibinfo {title} {Stochastic
  blockmodels and community structure in networks},}\ }\href {\doibase
  10.1103/PhysRevE.83.016107} {\bibfield  {journal} {\bibinfo  {journal}
  {Physical Review E}\ }\textbf {\bibinfo {volume} {83}},\ \bibinfo {pages}
  {016107} (\bibinfo {year} {2011})}\BibitemShut {NoStop}%
\bibitem [{\citenamefont {Newman}(2012)}]{newman_communities_2012}%
  \BibitemOpen
  \bibfield  {author} {\bibinfo {author} {\bibfnamefont {M.~E.~J.}\
  \bibnamefont {Newman}},\ }\bibfield  {title} {{\selectlanguage
  {english}\enquote {\bibinfo {title} {Communities, modules and large-scale
  structure in networks},}\ }}\href {\doibase 10.1038/nphys2162} {\bibfield
  {journal} {\bibinfo  {journal} {Nature Physics}\ }\textbf {\bibinfo {volume}
  {8}},\ \bibinfo {pages} {25--31} (\bibinfo {year} {2012})}\BibitemShut
  {NoStop}%
\bibitem [{\citenamefont {Porter}\ \emph {et~al.}(2009)\citenamefont {Porter},
  \citenamefont {Onnela},\ and\ \citenamefont
  {Mucha}}]{porter_communities_2009}%
  \BibitemOpen
  \bibfield  {author} {\bibinfo {author} {\bibfnamefont {Mason~A}\ \bibnamefont
  {Porter}}, \bibinfo {author} {\bibfnamefont {Jukka-Pekka}\ \bibnamefont
  {Onnela}}, \ and\ \bibinfo {author} {\bibfnamefont {Peter~J}\ \bibnamefont
  {Mucha}},\ }\bibfield  {title} {\enquote {\bibinfo {title} {Communities in
  {Networks}},}\ }\href {http://arxiv.org/abs/0902.3788} {\bibfield  {journal}
  {\bibinfo  {journal} {0902.3788}\ } (\bibinfo {year} {2009})},\ \bibinfo
  {note} {notices of the American Mathematical Society, Vol. 56, No. 9:
  1082-1097, 1164-1166, 2009}\BibitemShut {NoStop}%
\bibitem [{\citenamefont {Guimerà}\ and\ \citenamefont
  {Nunes~Amaral}(2005)}]{guimera_functional_2005}%
  \BibitemOpen
  \bibfield  {author} {\bibinfo {author} {\bibfnamefont {Roger}\ \bibnamefont
  {Guimerà}}\ and\ \bibinfo {author} {\bibfnamefont {Luís~A.}\ \bibnamefont
  {Nunes~Amaral}},\ }\bibfield  {title} {{\selectlanguage {english}\enquote
  {\bibinfo {title} {Functional cartography of complex metabolic networks},}\
  }}\href {\doibase 10.1038/nature03288} {\bibfield  {journal} {\bibinfo
  {journal} {Nature}\ }\textbf {\bibinfo {volume} {433}},\ \bibinfo {pages}
  {895--900} (\bibinfo {year} {2005})}\BibitemShut {NoStop}%
\bibitem [{\citenamefont {Ravasz}\ \emph {et~al.}(2002)\citenamefont {Ravasz},
  \citenamefont {Somera}, \citenamefont {Mongru}, \citenamefont {Oltvai},\ and\
  \citenamefont {Barabási}}]{ravasz_hierarchical_2002}%
  \BibitemOpen
  \bibfield  {author} {\bibinfo {author} {\bibfnamefont {E.}~\bibnamefont
  {Ravasz}}, \bibinfo {author} {\bibfnamefont {A.~L.}\ \bibnamefont {Somera}},
  \bibinfo {author} {\bibfnamefont {D.~A.}\ \bibnamefont {Mongru}}, \bibinfo
  {author} {\bibfnamefont {Z.~N.}\ \bibnamefont {Oltvai}}, \ and\ \bibinfo
  {author} {\bibfnamefont {A.-L.}\ \bibnamefont {Barabási}},\ }\bibfield
  {title} {{\selectlanguage {english}\enquote {\bibinfo {title} {Hierarchical
  {Organization} of {Modularity} in {Metabolic} {Networks}},}\ }}\href
  {\doibase 10.1126/science.1073374} {\bibfield  {journal} {\bibinfo  {journal}
  {Science}\ }\textbf {\bibinfo {volume} {297}},\ \bibinfo {pages} {1551--1555}
  (\bibinfo {year} {2002})}\BibitemShut {NoStop}%
\bibitem [{\citenamefont {Holme}\ \emph {et~al.}(2003)\citenamefont {Holme},
  \citenamefont {Huss},\ and\ \citenamefont {Jeong}}]{holme_subnetwork_2003}%
  \BibitemOpen
  \bibfield  {author} {\bibinfo {author} {\bibfnamefont {Petter}\ \bibnamefont
  {Holme}}, \bibinfo {author} {\bibfnamefont {Mikael}\ \bibnamefont {Huss}}, \
  and\ \bibinfo {author} {\bibfnamefont {Hawoong}\ \bibnamefont {Jeong}},\
  }\bibfield  {title} {{\selectlanguage {english}\enquote {\bibinfo {title}
  {Subnetwork hierarchies of biochemical pathways},}\ }}\href {\doibase
  10.1093/bioinformatics/btg033} {\bibfield  {journal} {\bibinfo  {journal}
  {Bioinformatics}\ }\textbf {\bibinfo {volume} {19}},\ \bibinfo {pages}
  {532--538} (\bibinfo {year} {2003})},\ \bibinfo {note} {publisher: Oxford
  Academic}\BibitemShut {NoStop}%
\bibitem [{\citenamefont {Guimerà}\ \emph {et~al.}(2004)\citenamefont
  {Guimerà}, \citenamefont {Sales-Pardo},\ and\ \citenamefont
  {Amaral}}]{guimera_modularity_2004}%
  \BibitemOpen
  \bibfield  {author} {\bibinfo {author} {\bibfnamefont {Roger}\ \bibnamefont
  {Guimerà}}, \bibinfo {author} {\bibfnamefont {Marta}\ \bibnamefont
  {Sales-Pardo}}, \ and\ \bibinfo {author} {\bibfnamefont {Luís A.~Nunes}\
  \bibnamefont {Amaral}},\ }\bibfield  {title} {\enquote {\bibinfo {title}
  {Modularity from fluctuations in random graphs and complex networks},}\
  }\href {\doibase 10.1103/PhysRevE.70.025101} {\bibfield  {journal} {\bibinfo
  {journal} {Physical Review E}\ }\textbf {\bibinfo {volume} {70}},\ \bibinfo
  {pages} {025101} (\bibinfo {year} {2004})}\BibitemShut {NoStop}%
\bibitem [{\citenamefont {McDiarmid}\ and\ \citenamefont
  {Skerman}(2016)}]{mcdiarmid_modularity_2016}%
  \BibitemOpen
  \bibfield  {author} {\bibinfo {author} {\bibfnamefont {Colin}\ \bibnamefont
  {McDiarmid}}\ and\ \bibinfo {author} {\bibfnamefont {Fiona}\ \bibnamefont
  {Skerman}},\ }\bibfield  {title} {\enquote {\bibinfo {title} {Modularity of
  tree-like and random regular graphs},}\ }\href
  {http://adsabs.harvard.edu/abs/2016arXiv160609101M} {\bibfield  {journal}
  {\bibinfo  {journal} {ArXiv e-prints}\ }\textbf {\bibinfo {volume} {1606}},\
  \bibinfo {pages} {arXiv:1606.09101} (\bibinfo {year} {2016})}\BibitemShut
  {NoStop}%
\bibitem [{\citenamefont {Bagrow}(2012)}]{bagrow_communities_2012}%
  \BibitemOpen
  \bibfield  {author} {\bibinfo {author} {\bibfnamefont {James~P.}\
  \bibnamefont {Bagrow}},\ }\bibfield  {title} {\enquote {\bibinfo {title}
  {Communities and bottlenecks: {Trees} and treelike networks have high
  modularity},}\ }\href {\doibase 10.1103/PhysRevE.85.066118} {\bibfield
  {journal} {\bibinfo  {journal} {Physical Review E}\ }\textbf {\bibinfo
  {volume} {85}},\ \bibinfo {pages} {066118} (\bibinfo {year}
  {2012})}\BibitemShut {NoStop}%
\bibitem [{\citenamefont {McDiarmid}\ and\ \citenamefont
  {Skerman}(2013)}]{mcdiarmid_modularity_2013}%
  \BibitemOpen
  \bibfield  {author} {\bibinfo {author} {\bibfnamefont {Colin}\ \bibnamefont
  {McDiarmid}}\ and\ \bibinfo {author} {\bibfnamefont {Fiona}\ \bibnamefont
  {Skerman}},\ }\bibfield  {title} {\enquote {\bibinfo {title} {Modularity in
  random regular graphs and lattices},}\ }\href {\doibase
  10.1016/j.endm.2013.07.063} {\bibfield  {journal} {\bibinfo  {journal}
  {Electronic Notes in Discrete Mathematics}\ }\textbf {\bibinfo {volume}
  {43}},\ \bibinfo {pages} {431--437} (\bibinfo {year} {2013})}\BibitemShut
  {NoStop}%
\bibitem [{\citenamefont {Peixoto}(2019)}]{peixoto_bayesian_2019}%
  \BibitemOpen
  \bibfield  {author} {\bibinfo {author} {\bibfnamefont {Tiago~P.}\
  \bibnamefont {Peixoto}},\ }\bibfield  {title} {{\selectlanguage
  {english}\enquote {\bibinfo {title} {Bayesian {Stochastic}
  {Blockmodeling}},}\ }}in\ \href {\doibase 10.1002/9781119483298.ch11}
  {{\selectlanguage {english}\emph {\bibinfo {booktitle} {Advances in {Network}
  {Clustering} and {Blockmodeling}}}}}\ (\bibinfo  {publisher} {John Wiley \&
  Sons, Ltd},\ \bibinfo {year} {2019})\ pp.\ \bibinfo {pages}
  {289--332}\BibitemShut {NoStop}%
\bibitem [{\citenamefont {Bui}\ \emph {et~al.}(1987)\citenamefont {Bui},
  \citenamefont {Chaudhuri}, \citenamefont {Leighton},\ and\ \citenamefont
  {Sipser}}]{bui_graph_1987}%
  \BibitemOpen
  \bibfield  {author} {\bibinfo {author} {\bibfnamefont {T.~N.}\ \bibnamefont
  {Bui}}, \bibinfo {author} {\bibfnamefont {S.}~\bibnamefont {Chaudhuri}},
  \bibinfo {author} {\bibfnamefont {F.~T.}\ \bibnamefont {Leighton}}, \ and\
  \bibinfo {author} {\bibfnamefont {M.}~\bibnamefont {Sipser}},\ }\bibfield
  {title} {{\selectlanguage {english}\enquote {\bibinfo {title} {Graph
  bisection algorithms with good average case behavior},}\ }}\href {\doibase
  10.1007/BF02579448} {\bibfield  {journal} {\bibinfo  {journal}
  {Combinatorica}\ }\textbf {\bibinfo {volume} {7}},\ \bibinfo {pages}
  {171--191} (\bibinfo {year} {1987})}\BibitemShut {NoStop}%
\bibitem [{\citenamefont {Dyer}\ and\ \citenamefont
  {Frieze}(1989)}]{dyer_solution_1989}%
  \BibitemOpen
  \bibfield  {author} {\bibinfo {author} {\bibfnamefont {M.~E}\ \bibnamefont
  {Dyer}}\ and\ \bibinfo {author} {\bibfnamefont {A.~M}\ \bibnamefont
  {Frieze}},\ }\bibfield  {title} {\enquote {\bibinfo {title} {The solution of
  some random {NP}-hard problems in polynomial expected time},}\ }\href
  {\doibase 10.1016/0196-6774(89)90001-1} {\bibfield  {journal} {\bibinfo
  {journal} {Journal of Algorithms}\ }\textbf {\bibinfo {volume} {10}},\
  \bibinfo {pages} {451--489} (\bibinfo {year} {1989})}\BibitemShut {NoStop}%
\bibitem [{\citenamefont {Condon}\ and\ \citenamefont
  {Karp}(2001)}]{condon_algorithms_2001}%
  \BibitemOpen
  \bibfield  {author} {\bibinfo {author} {\bibfnamefont {Anne}\ \bibnamefont
  {Condon}}\ and\ \bibinfo {author} {\bibfnamefont {Richard~M.}\ \bibnamefont
  {Karp}},\ }\bibfield  {title} {{\selectlanguage {english}\enquote {\bibinfo
  {title} {Algorithms for graph partitioning on the planted partition model},}\
  }}\href {\doibase 10.1002/1098-2418(200103)18:2<116::AID-RSA1001>3.0.CO;2-2}
  {\bibfield  {journal} {\bibinfo  {journal} {Random Structures \& Algorithms}\
  }\textbf {\bibinfo {volume} {18}},\ \bibinfo {pages} {116--140} (\bibinfo
  {year} {2001})}\BibitemShut {NoStop}%
\bibitem [{\citenamefont {Fortunato}\ and\ \citenamefont
  {Barthélemy}(2007)}]{fortunato_resolution_2007}%
  \BibitemOpen
  \bibfield  {author} {\bibinfo {author} {\bibfnamefont {Santo}\ \bibnamefont
  {Fortunato}}\ and\ \bibinfo {author} {\bibfnamefont {Marc}\ \bibnamefont
  {Barthélemy}},\ }\bibfield  {title} {{\selectlanguage {english}\enquote
  {\bibinfo {title} {Resolution limit in community detection},}\ }}\href
  {\doibase 10.1073/pnas.0605965104} {\bibfield  {journal} {\bibinfo  {journal}
  {Proceedings of the National Academy of Sciences}\ }\textbf {\bibinfo
  {volume} {104}},\ \bibinfo {pages} {36--41} (\bibinfo {year}
  {2007})}\BibitemShut {NoStop}%
\bibitem [{\citenamefont {Zhang}\ and\ \citenamefont
  {Moore}(2014)}]{zhang_scalable_2014}%
  \BibitemOpen
  \bibfield  {author} {\bibinfo {author} {\bibfnamefont {Pan}\ \bibnamefont
  {Zhang}}\ and\ \bibinfo {author} {\bibfnamefont {Cristopher}\ \bibnamefont
  {Moore}},\ }\bibfield  {title} {{\selectlanguage {english}\enquote {\bibinfo
  {title} {Scalable detection of statistically significant communities and
  hierarchies, using message passing for modularity},}\ }}\href {\doibase
  10.1073/pnas.1409770111} {\bibfield  {journal} {\bibinfo  {journal}
  {Proceedings of the National Academy of Sciences}\ }\textbf {\bibinfo
  {volume} {111}},\ \bibinfo {pages} {18144--18149} (\bibinfo {year} {2014})},\
  \bibinfo {note} {publisher: National Academy of Sciences Section: Physical
  Sciences}\BibitemShut {NoStop}%
\bibitem [{\citenamefont {Newman}(2016)}]{newman_equivalence_2016}%
  \BibitemOpen
  \bibfield  {author} {\bibinfo {author} {\bibfnamefont {M.~E.~J.}\
  \bibnamefont {Newman}},\ }\bibfield  {title} {\enquote {\bibinfo {title}
  {Equivalence between modularity optimization and maximum likelihood methods
  for community detection},}\ }\href {\doibase 10.1103/PhysRevE.94.052315}
  {\bibfield  {journal} {\bibinfo  {journal} {Physical Review E}\ }\textbf
  {\bibinfo {volume} {94}} (\bibinfo {year} {2016}),\
  10.1103/PhysRevE.94.052315}\BibitemShut {NoStop}%
\bibitem [{\citenamefont {Blondel}\ \emph {et~al.}(2008)\citenamefont
  {Blondel}, \citenamefont {Guillaume}, \citenamefont {Lambiotte},\ and\
  \citenamefont {Lefebvre}}]{blondel_fast_2008}%
  \BibitemOpen
  \bibfield  {author} {\bibinfo {author} {\bibfnamefont {Vincent~D.}\
  \bibnamefont {Blondel}}, \bibinfo {author} {\bibfnamefont {Jean-Loup}\
  \bibnamefont {Guillaume}}, \bibinfo {author} {\bibfnamefont {Renaud}\
  \bibnamefont {Lambiotte}}, \ and\ \bibinfo {author} {\bibfnamefont {Etienne}\
  \bibnamefont {Lefebvre}},\ }\bibfield  {title} {{\selectlanguage
  {english}\enquote {\bibinfo {title} {Fast unfolding of communities in large
  networks},}\ }}\href {\doibase 10.1088/1742-5468/2008/10/P10008} {\bibfield
  {journal} {\bibinfo  {journal} {Journal of Statistical Mechanics: Theory and
  Experiment}\ }\textbf {\bibinfo {volume} {2008}},\ \bibinfo {pages} {P10008}
  (\bibinfo {year} {2008})}\BibitemShut {NoStop}%
\bibitem [{\citenamefont
  {Peixoto}(2020{\natexlab{a}})}]{peixoto_merge-split_2020}%
  \BibitemOpen
  \bibfield  {author} {\bibinfo {author} {\bibfnamefont {Tiago~P.}\
  \bibnamefont {Peixoto}},\ }\bibfield  {title} {\enquote {\bibinfo {title}
  {Merge-split {Markov} chain {Monte} {Carlo} for community detection},}\
  }\href {http://arxiv.org/abs/2003.07070} {\bibfield  {journal} {\bibinfo
  {journal} {arXiv:2003.07070 [physics, stat]}\ } (\bibinfo {year}
  {2020}{\natexlab{a}})},\ \bibinfo {note} {arXiv: 2003.07070}\BibitemShut
  {NoStop}%
\bibitem [{\citenamefont {Peixoto}(2017)}]{peixoto_nonparametric_2017}%
  \BibitemOpen
  \bibfield  {author} {\bibinfo {author} {\bibfnamefont {Tiago~P.}\
  \bibnamefont {Peixoto}},\ }\bibfield  {title} {\enquote {\bibinfo {title}
  {Nonparametric {Bayesian} inference of the microcanonical stochastic block
  model},}\ }\href {\doibase 10.1103/PhysRevE.95.012317} {\bibfield  {journal}
  {\bibinfo  {journal} {Physical Review E}\ }\textbf {\bibinfo {volume} {95}},\
  \bibinfo {pages} {012317} (\bibinfo {year} {2017})}\BibitemShut {NoStop}%
\bibitem [{\citenamefont {Newman}(2006)}]{newman_modularity_2006}%
  \BibitemOpen
  \bibfield  {author} {\bibinfo {author} {\bibfnamefont {M.~E.~J.}\
  \bibnamefont {Newman}},\ }\bibfield  {title} {{\selectlanguage
  {english}\enquote {\bibinfo {title} {Modularity and community structure in
  networks},}\ }}\href {\doibase 10.1073/pnas.0601602103} {\bibfield  {journal}
  {\bibinfo  {journal} {Proceedings of the National Academy of Sciences}\
  }\textbf {\bibinfo {volume} {103}},\ \bibinfo {pages} {8577--8582} (\bibinfo
  {year} {2006})}\BibitemShut {NoStop}%
\bibitem [{\citenamefont {Reichardt}\ and\ \citenamefont
  {Bornholdt}(2006)}]{reichardt_statistical_2006}%
  \BibitemOpen
  \bibfield  {author} {\bibinfo {author} {\bibfnamefont {Jörg}\ \bibnamefont
  {Reichardt}}\ and\ \bibinfo {author} {\bibfnamefont {Stefan}\ \bibnamefont
  {Bornholdt}},\ }\bibfield  {title} {\enquote {\bibinfo {title} {Statistical
  mechanics of community detection},}\ }\href {\doibase
  10.1103/PhysRevE.74.016110} {\bibfield  {journal} {\bibinfo  {journal}
  {Physical Review E}\ }\textbf {\bibinfo {volume} {74}},\ \bibinfo {pages}
  {016110} (\bibinfo {year} {2006})}\BibitemShut {NoStop}%
\bibitem [{\citenamefont {McDiarmid}\ and\ \citenamefont
  {Skerman}()}]{mcdiarmid_modularity_nodate}%
  \BibitemOpen
  \bibfield  {author} {\bibinfo {author} {\bibfnamefont {Colin}\ \bibnamefont
  {McDiarmid}}\ and\ \bibinfo {author} {\bibfnamefont {Fiona}\ \bibnamefont
  {Skerman}},\ }\bibfield  {title} {{\selectlanguage {english}\enquote
  {\bibinfo {title} {Modularity of regular and treelike graphs},}\ }}\href
  {\doibase 10.1093/comnet/cnx046} {\bibfield  {journal} {\bibinfo  {journal}
  {Journal of Complex Networks}\ }10.1093/comnet/cnx046}\BibitemShut {NoStop}%
\bibitem [{\citenamefont {Good}\ \emph {et~al.}(2010)\citenamefont {Good},
  \citenamefont {de~Montjoye},\ and\ \citenamefont
  {Clauset}}]{good_performance_2010}%
  \BibitemOpen
  \bibfield  {author} {\bibinfo {author} {\bibfnamefont {Benjamin~H.}\
  \bibnamefont {Good}}, \bibinfo {author} {\bibfnamefont {Yves-Alexandre}\
  \bibnamefont {de~Montjoye}}, \ and\ \bibinfo {author} {\bibfnamefont {Aaron}\
  \bibnamefont {Clauset}},\ }\bibfield  {title} {\enquote {\bibinfo {title}
  {Performance of modularity maximization in practical contexts},}\ }\href
  {\doibase 10.1103/PhysRevE.81.046106} {\bibfield  {journal} {\bibinfo
  {journal} {Physical Review E}\ }\textbf {\bibinfo {volume} {81}},\ \bibinfo
  {pages} {046106} (\bibinfo {year} {2010})}\BibitemShut {NoStop}%
\bibitem [{\citenamefont {Ghasemian}\ \emph {et~al.}(2019)\citenamefont
  {Ghasemian}, \citenamefont {Hosseinmardi},\ and\ \citenamefont
  {Clauset}}]{ghasemian_evaluating_2019}%
  \BibitemOpen
  \bibfield  {author} {\bibinfo {author} {\bibfnamefont {Amir}\ \bibnamefont
  {Ghasemian}}, \bibinfo {author} {\bibfnamefont {Homa}\ \bibnamefont
  {Hosseinmardi}}, \ and\ \bibinfo {author} {\bibfnamefont {Aaron}\
  \bibnamefont {Clauset}},\ }\bibfield  {title} {\enquote {\bibinfo {title}
  {Evaluating {Overfit} and {Underfit} in {Models} of {Network} {Community}
  {Structure}},}\ }\href {\doibase 10.1109/TKDE.2019.2911585} {\bibfield
  {journal} {\bibinfo  {journal} {IEEE Transactions on Knowledge and Data
  Engineering}\ ,\ \bibinfo {pages} {1--1}} (\bibinfo {year}
  {2019})}\BibitemShut {NoStop}%
\bibitem [{\citenamefont {Grünwald}(2007)}]{grunwald_minimum_2007}%
  \BibitemOpen
  \bibfield  {author} {\bibinfo {author} {\bibfnamefont {Peter~D.}\
  \bibnamefont {Grünwald}},\ }\href@noop {} {\emph {\bibinfo {title} {The
  {Minimum} {Description} {Length} {Principle}}}}\ (\bibinfo  {publisher} {The
  MIT Press},\ \bibinfo {year} {2007})\BibitemShut {NoStop}%
\bibitem [{\citenamefont {Metropolis}\ \emph {et~al.}(1953)\citenamefont
  {Metropolis}, \citenamefont {Rosenbluth}, \citenamefont {Rosenbluth},
  \citenamefont {Teller},\ and\ \citenamefont
  {Teller}}]{metropolis_equation_1953}%
  \BibitemOpen
  \bibfield  {author} {\bibinfo {author} {\bibfnamefont {Nicholas}\
  \bibnamefont {Metropolis}}, \bibinfo {author} {\bibfnamefont {Arianna~W.}\
  \bibnamefont {Rosenbluth}}, \bibinfo {author} {\bibfnamefont {Marshall~N.}\
  \bibnamefont {Rosenbluth}}, \bibinfo {author} {\bibfnamefont {Augusta~H.}\
  \bibnamefont {Teller}}, \ and\ \bibinfo {author} {\bibfnamefont {Edward}\
  \bibnamefont {Teller}},\ }\bibfield  {title} {\enquote {\bibinfo {title}
  {Equation of {State} {Calculations} by {Fast} {Computing} {Machines}},}\
  }\href {\doibase 10.1063/1.1699114} {\bibfield  {journal} {\bibinfo
  {journal} {The Journal of Chemical Physics}\ }\textbf {\bibinfo {volume}
  {21}},\ \bibinfo {pages} {1087} (\bibinfo {year} {1953})}\BibitemShut
  {NoStop}%
\bibitem [{\citenamefont {Hastings}(1970)}]{hastings_monte_1970}%
  \BibitemOpen
  \bibfield  {author} {\bibinfo {author} {\bibfnamefont {W.~K.}\ \bibnamefont
  {Hastings}},\ }\bibfield  {title} {\enquote {\bibinfo {title} {Monte {Carlo}
  sampling methods using {Markov} chains and their applications},}\ }\href
  {\doibase 10.1093/biomet/57.1.97} {\bibfield  {journal} {\bibinfo  {journal}
  {Biometrika}\ }\textbf {\bibinfo {volume} {57}},\ \bibinfo {pages} {97 --109}
  (\bibinfo {year} {1970})}\BibitemShut {NoStop}%
\bibitem [{\citenamefont
  {Peixoto}(2014{\natexlab{a}})}]{peixoto_graph-tool_2014}%
  \BibitemOpen
  \bibfield  {author} {\bibinfo {author} {\bibfnamefont {Tiago~P.}\
  \bibnamefont {Peixoto}},\ }\bibfield  {title} {\enquote {\bibinfo {title}
  {The \texttt{graph-tool} python library},}\ }\href {\doibase
  10.6084/m9.figshare.1164194} {\bibfield  {journal} {\bibinfo  {journal}
  {figshare}\ } (\bibinfo {year} {2014}{\natexlab{a}}),\
  10.6084/m9.figshare.1164194},\ \bibinfo {note} {available at
  \url{https://graph-tool.skewed.de}.}\BibitemShut {Stop}%
\bibitem [{\citenamefont {Arenas}\ \emph {et~al.}(2008)\citenamefont {Arenas},
  \citenamefont {Fernández},\ and\ \citenamefont
  {Gómez}}]{arenas_analysis_2008}%
  \BibitemOpen
  \bibfield  {author} {\bibinfo {author} {\bibfnamefont {A.}~\bibnamefont
  {Arenas}}, \bibinfo {author} {\bibfnamefont {A.}~\bibnamefont {Fernández}},
  \ and\ \bibinfo {author} {\bibfnamefont {S.}~\bibnamefont {Gómez}},\
  }\bibfield  {title} {{\selectlanguage {english}\enquote {\bibinfo {title}
  {Analysis of the structure of complex networks at different resolution
  levels},}\ }}\href {\doibase 10.1088/1367-2630/10/5/053039} {\bibfield
  {journal} {\bibinfo  {journal} {New Journal of Physics}\ }\textbf {\bibinfo
  {volume} {10}},\ \bibinfo {pages} {053039} (\bibinfo {year} {2008})},\
  \bibinfo {note} {publisher: IOP Publishing}\BibitemShut {NoStop}%
\bibitem [{\citenamefont {Ronhovde}\ and\ \citenamefont
  {Nussinov}(2009)}]{ronhovde_multiresolution_2009}%
  \BibitemOpen
  \bibfield  {author} {\bibinfo {author} {\bibfnamefont {Peter}\ \bibnamefont
  {Ronhovde}}\ and\ \bibinfo {author} {\bibfnamefont {Zohar}\ \bibnamefont
  {Nussinov}},\ }\bibfield  {title} {\enquote {\bibinfo {title}
  {Multiresolution community detection for megascale networks by
  information-based replica correlations},}\ }\href {\doibase
  10.1103/PhysRevE.80.016109} {\bibfield  {journal} {\bibinfo  {journal}
  {Physical Review E}\ }\textbf {\bibinfo {volume} {80}},\ \bibinfo {pages}
  {016109} (\bibinfo {year} {2009})}\BibitemShut {NoStop}%
\bibitem [{\citenamefont {Lancichinetti}\ and\ \citenamefont
  {Fortunato}(2011)}]{lancichinetti_limits_2011}%
  \BibitemOpen
  \bibfield  {author} {\bibinfo {author} {\bibfnamefont {Andrea}\ \bibnamefont
  {Lancichinetti}}\ and\ \bibinfo {author} {\bibfnamefont {Santo}\ \bibnamefont
  {Fortunato}},\ }\bibfield  {title} {\enquote {\bibinfo {title} {Limits of
  modularity maximization in community detection},}\ }\href {\doibase
  10.1103/PhysRevE.84.066122} {\bibfield  {journal} {\bibinfo  {journal}
  {Physical Review E}\ }\textbf {\bibinfo {volume} {84}},\ \bibinfo {pages}
  {066122} (\bibinfo {year} {2011})}\BibitemShut {NoStop}%
\bibitem [{\citenamefont {Ronhovde}\ and\ \citenamefont
  {Nussinov}(2015)}]{ronhovde_local_2015}%
  \BibitemOpen
  \bibfield  {author} {\bibinfo {author} {\bibfnamefont {Peter}\ \bibnamefont
  {Ronhovde}}\ and\ \bibinfo {author} {\bibfnamefont {Zohar}\ \bibnamefont
  {Nussinov}},\ }\bibfield  {title} {{\selectlanguage {english}\enquote
  {\bibinfo {title} {Local multiresolution order in community detection},}\
  }}\href {\doibase 10.1088/1742-5468/2015/01/P01001} {\bibfield  {journal}
  {\bibinfo  {journal} {Journal of Statistical Mechanics: Theory and
  Experiment}\ }\textbf {\bibinfo {volume} {2015}},\ \bibinfo {pages} {P01001}
  (\bibinfo {year} {2015})},\ \bibinfo {note} {publisher: IOP
  Publishing}\BibitemShut {NoStop}%
\bibitem [{\citenamefont {Decelle}\ \emph {et~al.}(2011)\citenamefont
  {Decelle}, \citenamefont {Krzakala}, \citenamefont {Moore},\ and\
  \citenamefont {Zdeborová}}]{decelle_asymptotic_2011}%
  \BibitemOpen
  \bibfield  {author} {\bibinfo {author} {\bibfnamefont {Aurelien}\
  \bibnamefont {Decelle}}, \bibinfo {author} {\bibfnamefont {Florent}\
  \bibnamefont {Krzakala}}, \bibinfo {author} {\bibfnamefont {Cristopher}\
  \bibnamefont {Moore}}, \ and\ \bibinfo {author} {\bibfnamefont {Lenka}\
  \bibnamefont {Zdeborová}},\ }\bibfield  {title} {\enquote {\bibinfo {title}
  {Asymptotic analysis of the stochastic block model for modular networks and
  its algorithmic applications},}\ }\href {\doibase 10.1103/PhysRevE.84.066106}
  {\bibfield  {journal} {\bibinfo  {journal} {Physical Review E}\ }\textbf
  {\bibinfo {volume} {84}},\ \bibinfo {pages} {066106} (\bibinfo {year}
  {2011})}\BibitemShut {NoStop}%
\bibitem [{\citenamefont {Peixoto}(2013)}]{peixoto_parsimonious_2013}%
  \BibitemOpen
  \bibfield  {author} {\bibinfo {author} {\bibfnamefont {Tiago~P.}\
  \bibnamefont {Peixoto}},\ }\bibfield  {title} {\enquote {\bibinfo {title}
  {Parsimonious {Module} {Inference} in {Large} {Networks}},}\ }\href {\doibase
  10.1103/PhysRevLett.110.148701} {\bibfield  {journal} {\bibinfo  {journal}
  {Physical Review Letters}\ }\textbf {\bibinfo {volume} {110}},\ \bibinfo
  {pages} {148701} (\bibinfo {year} {2013})}\BibitemShut {NoStop}%
\bibitem [{\citenamefont
  {Peixoto}(2014{\natexlab{b}})}]{peixoto_hierarchical_2014}%
  \BibitemOpen
  \bibfield  {author} {\bibinfo {author} {\bibfnamefont {Tiago~P.}\
  \bibnamefont {Peixoto}},\ }\bibfield  {title} {\enquote {\bibinfo {title}
  {Hierarchical {Block} {Structures} and {High}-{Resolution} {Model}
  {Selection} in {Large} {Networks}},}\ }\href {\doibase
  10.1103/PhysRevX.4.011047} {\bibfield  {journal} {\bibinfo  {journal}
  {Physical Review X}\ }\textbf {\bibinfo {volume} {4}},\ \bibinfo {pages}
  {011047} (\bibinfo {year} {2014}{\natexlab{b}})}\BibitemShut {NoStop}%
\bibitem [{\citenamefont {Kunegis}(2013)}]{kunegis_konect:_2013}%
  \BibitemOpen
  \bibfield  {author} {\bibinfo {author} {\bibfnamefont {Jérôme}\
  \bibnamefont {Kunegis}},\ }\bibfield  {title} {\enquote {\bibinfo {title}
  {{KONECT}: {The} {Koblenz} {Network} {Collection}},}\ }in\ \href {\doibase
  10.1145/2487788.2488173} {\emph {\bibinfo {booktitle} {Proceedings of the
  {22Nd} {International} {Conference} on {World} {Wide} {Web}}}},\ \bibinfo
  {series and number} {{WWW} '13 {Companion}}\ (\bibinfo  {publisher} {ACM},\
  \bibinfo {address} {New York, NY, USA},\ \bibinfo {year} {2013})\ pp.\
  \bibinfo {pages} {1343--1350}\BibitemShut {NoStop}%
\bibitem [{\citenamefont {Girvan}\ and\ \citenamefont
  {Newman}(2002)}]{girvan_community_2002}%
  \BibitemOpen
  \bibfield  {author} {\bibinfo {author} {\bibfnamefont {M.}~\bibnamefont
  {Girvan}}\ and\ \bibinfo {author} {\bibfnamefont {M.~E.~J.}\ \bibnamefont
  {Newman}},\ }\bibfield  {title} {\enquote {\bibinfo {title} {Community
  structure in social and biological networks},}\ }\href {\doibase
  10.1073/pnas.122653799} {\bibfield  {journal} {\bibinfo  {journal}
  {Proceedings of the National Academy of Sciences}\ }\textbf {\bibinfo
  {volume} {99}},\ \bibinfo {pages} {7821 --7826} (\bibinfo {year}
  {2002})}\BibitemShut {NoStop}%
\bibitem [{\citenamefont {Krebs}()}]{krebs_political_nodate}%
  \BibitemOpen
  \bibfield  {author} {\bibinfo {author} {\bibfnamefont {V}~\bibnamefont
  {Krebs}},\ }\bibfield  {title} {\enquote {\bibinfo {title} {Political {Books}
  {Network}},}\ }\href@noop {} {\bibinfo  {journal} {unpublished, retrieved
  from Mark Newman's website:
  \url{http://www-personal.umich.edu/{\textasciitilde}mejn/netdata/}}\
  }\BibitemShut {NoStop}%
\bibitem [{\citenamefont {Harris}\ \emph {et~al.}(2009)\citenamefont {Harris},
  \citenamefont {Halpern}, \citenamefont {Whitsel}, \citenamefont {Hussey},
  \citenamefont {Tabor}, \citenamefont {Entzel},\ and\ \citenamefont
  {Udry}}]{harris_national_2009}%
  \BibitemOpen
\bibfield  {journal} {  }\bibfield  {author} {\bibinfo {author} {\bibfnamefont
  {Kathleen~Mullan}\ \bibnamefont {Harris}}, \bibinfo {author} {\bibfnamefont
  {Carolyn~T.}\ \bibnamefont {Halpern}}, \bibinfo {author} {\bibfnamefont
  {Eric}\ \bibnamefont {Whitsel}}, \bibinfo {author} {\bibfnamefont {Jon}\
  \bibnamefont {Hussey}}, \bibinfo {author} {\bibfnamefont {Joyce}\
  \bibnamefont {Tabor}}, \bibinfo {author} {\bibfnamefont {Pamela}\
  \bibnamefont {Entzel}}, \ and\ \bibinfo {author} {\bibfnamefont {J.~Richard}\
  \bibnamefont {Udry}},\ }\bibfield  {title} {\enquote {\bibinfo {title} {The
  national longitudinal study of adolescent to adult health: {Research}
  design},}\ }\href@noop {} {\bibfield  {journal} {\bibinfo  {journal} {See
  http://www. cpc. unc. edu/projects/addhealth/design (accessed 9 April 2015)}\
  } (\bibinfo {year} {2009})}\BibitemShut {NoStop}%
\bibitem [{\citenamefont
  {Peixoto}(2020{\natexlab{b}})}]{peixoto_revealing_2020}%
  \BibitemOpen
  \bibfield  {author} {\bibinfo {author} {\bibfnamefont {Tiago~P.}\
  \bibnamefont {Peixoto}},\ }\bibfield  {title} {\enquote {\bibinfo {title}
  {Revealing consensus and dissensus between network partitions},}\ }\href
  {http://arxiv.org/abs/2005.13977} {\bibfield  {journal} {\bibinfo  {journal}
  {arXiv:2005.13977 [physics, stat]}\ } (\bibinfo {year}
  {2020}{\natexlab{b}})},\ \bibinfo {note} {arXiv: 2005.13977}\BibitemShut
  {NoStop}%
\bibitem [{\citenamefont {Krebs}(2002)}]{krebs_uncloaking_2002}%
  \BibitemOpen
  \bibfield  {author} {\bibinfo {author} {\bibfnamefont {Valdis}\ \bibnamefont
  {Krebs}},\ }\bibfield  {title} {{\selectlanguage {english}\enquote {\bibinfo
  {title} {Uncloaking {Terrorist} {Networks}},}\ }}\href {\doibase
  10.5210/fm.v7i4.941} {\bibfield  {journal} {\bibinfo  {journal} {First
  Monday}\ }\textbf {\bibinfo {volume} {7}} (\bibinfo {year} {2002}),\
  10.5210/fm.v7i4.941}\BibitemShut {NoStop}%
\bibitem [{\citenamefont {Lusseau}\ \emph {et~al.}(2003)\citenamefont
  {Lusseau}, \citenamefont {Schneider}, \citenamefont {Boisseau}, \citenamefont
  {Haase}, \citenamefont {Slooten},\ and\ \citenamefont
  {Dawson}}]{lusseau_bottlenose_2003}%
  \BibitemOpen
  \bibfield  {author} {\bibinfo {author} {\bibfnamefont {David}\ \bibnamefont
  {Lusseau}}, \bibinfo {author} {\bibfnamefont {Karsten}\ \bibnamefont
  {Schneider}}, \bibinfo {author} {\bibfnamefont {Oliver~J.}\ \bibnamefont
  {Boisseau}}, \bibinfo {author} {\bibfnamefont {Patti}\ \bibnamefont {Haase}},
  \bibinfo {author} {\bibfnamefont {Elisabeth}\ \bibnamefont {Slooten}}, \ and\
  \bibinfo {author} {\bibfnamefont {Steve~M.}\ \bibnamefont {Dawson}},\
  }\bibfield  {title} {{\selectlanguage {english}\enquote {\bibinfo {title}
  {The bottlenose dolphin community of {Doubtful} {Sound} features a large
  proportion of long-lasting associations},}\ }}\href {\doibase
  10.1007/s00265-003-0651-y} {\bibfield  {journal} {\bibinfo  {journal}
  {Behavioral Ecology and Sociobiology}\ }\textbf {\bibinfo {volume} {54}},\
  \bibinfo {pages} {396--405} (\bibinfo {year} {2003})}\BibitemShut {NoStop}%
\bibitem [{\citenamefont {Reza}\ and\ \citenamefont
  {Huan}(2009)}]{reza_social_2009}%
  \BibitemOpen
  \bibfield  {author} {\bibinfo {author} {\bibfnamefont {Zafarani}\
  \bibnamefont {Reza}}\ and\ \bibinfo {author} {\bibfnamefont {Liu}\
  \bibnamefont {Huan}},\ }\bibfield  {title} {\enquote {\bibinfo {title}
  {Social computing data repository},}\ }\href@noop {} {\  (\bibinfo {year}
  {2009})}\BibitemShut {NoStop}%
\bibitem [{\citenamefont {Adamic}\ and\ \citenamefont
  {Glance}(2005)}]{adamic_political_2005}%
  \BibitemOpen
  \bibfield  {author} {\bibinfo {author} {\bibfnamefont {Lada~A.}\ \bibnamefont
  {Adamic}}\ and\ \bibinfo {author} {\bibfnamefont {Natalie}\ \bibnamefont
  {Glance}},\ }\bibfield  {title} {\enquote {\bibinfo {title} {The political
  blogosphere and the 2004 {U}.{S}. election: divided they blog},}\ }in\ \href
  {\doibase 10.1145/1134271.1134277} {\emph {\bibinfo {booktitle} {Proceedings
  of the 3rd international workshop on {Link} discovery}}},\ \bibinfo {series
  and number} {{LinkKDD} '05}\ (\bibinfo  {publisher} {ACM},\ \bibinfo
  {address} {New York, NY, USA},\ \bibinfo {year} {2005})\ pp.\ \bibinfo
  {pages} {36--43}\BibitemShut {NoStop}%
\bibitem [{\citenamefont {Jeong}\ \emph {et~al.}(2001)\citenamefont {Jeong},
  \citenamefont {Mason}, \citenamefont {Barabási},\ and\ \citenamefont
  {Oltvai}}]{jeong_lethality_2001}%
  \BibitemOpen
  \bibfield  {author} {\bibinfo {author} {\bibfnamefont {H.}~\bibnamefont
  {Jeong}}, \bibinfo {author} {\bibfnamefont {S.~P.}\ \bibnamefont {Mason}},
  \bibinfo {author} {\bibfnamefont {A.-L.}\ \bibnamefont {Barabási}}, \ and\
  \bibinfo {author} {\bibfnamefont {Z.~N.}\ \bibnamefont {Oltvai}},\ }\bibfield
   {title} {{\selectlanguage {english}\enquote {\bibinfo {title} {Lethality and
  centrality in protein networks},}\ }}\href {\doibase 10.1038/35075138}
  {\bibfield  {journal} {\bibinfo  {journal} {Nature}\ }\textbf {\bibinfo
  {volume} {411}},\ \bibinfo {pages} {41--42} (\bibinfo {year} {2001})},\
  \bibinfo {note} {number: 6833 Publisher: Nature Publishing Group}\BibitemShut
  {NoStop}%
\bibitem [{\citenamefont {Fellbaum}(1998)}]{fellbaum_wordnet_1998}%
  \BibitemOpen
  \bibfield  {author} {\bibinfo {author} {\bibfnamefont {Christiane}\
  \bibnamefont {Fellbaum}},\ }\bibfield  {title} {\enquote {\bibinfo {title}
  {{WordNet}: {An} electronic lexical database {MIT} {Press}},}\ }\href@noop {}
  {\bibfield  {journal} {\bibinfo  {journal} {Cambridge, Massachusetts}\ }
  (\bibinfo {year} {1998})}\BibitemShut {NoStop}%
\bibitem [{\citenamefont {Leskovec}\ \emph {et~al.}(2007)\citenamefont
  {Leskovec}, \citenamefont {Kleinberg},\ and\ \citenamefont
  {Faloutsos}}]{leskovec_graph_2007}%
  \BibitemOpen
  \bibfield  {author} {\bibinfo {author} {\bibfnamefont {Jure}\ \bibnamefont
  {Leskovec}}, \bibinfo {author} {\bibfnamefont {Jon}\ \bibnamefont
  {Kleinberg}}, \ and\ \bibinfo {author} {\bibfnamefont {Christos}\
  \bibnamefont {Faloutsos}},\ }\bibfield  {title} {\enquote {\bibinfo {title}
  {Graph evolution: {Densification} and shrinking diameters},}\ }\href
  {\doibase 10.1145/1217299.1217301} {\bibfield  {journal} {\bibinfo  {journal}
  {ACM Trans. Knowl. Discov. Data}\ }\textbf {\bibinfo {volume} {1}} (\bibinfo
  {year} {2007}),\ 10.1145/1217299.1217301}\BibitemShut {NoStop}%
\bibitem [{\citenamefont {Decker}\ \emph {et~al.}(1991)\citenamefont {Decker},
  \citenamefont {Kohfeld}, \citenamefont {Rosenfeld},\ and\ \citenamefont
  {Sprague}}]{decker_st_1991}%
  \BibitemOpen
  \bibfield  {author} {\bibinfo {author} {\bibfnamefont {Scott}\ \bibnamefont
  {Decker}}, \bibinfo {author} {\bibfnamefont {Carol~W.}\ \bibnamefont
  {Kohfeld}}, \bibinfo {author} {\bibfnamefont {Richard}\ \bibnamefont
  {Rosenfeld}}, \ and\ \bibinfo {author} {\bibfnamefont {John}\ \bibnamefont
  {Sprague}},\ }\href@noop {} {\emph {\bibinfo {title} {St. {Louis} {Homicide}
  {Project}: {Local} {Responses} to a {National} {Problem}}}}\ (\bibinfo
  {publisher} {St. Louis, MO: University of Missouri-St. Louis},\ \bibinfo
  {year} {1991})\BibitemShut {NoStop}%
\bibitem [{\citenamefont {Yen}\ and\ \citenamefont
  {Larremore}(2020)}]{yen_community_2020}%
  \BibitemOpen
  \bibfield  {author} {\bibinfo {author} {\bibfnamefont {Tzu-Chi}\ \bibnamefont
  {Yen}}\ and\ \bibinfo {author} {\bibfnamefont {Daniel~B.}\ \bibnamefont
  {Larremore}},\ }\bibfield  {title} {\enquote {\bibinfo {title} {Community
  {Detection} in {Bipartite} {Networks} with {Stochastic} {Blockmodels}},}\
  }\href {http://arxiv.org/abs/2001.11818} {\bibfield  {journal} {\bibinfo
  {journal} {arXiv:2001.11818 [physics, stat]}\ } (\bibinfo {year} {2020})},\
  \bibinfo {note} {arXiv: 2001.11818}\BibitemShut {NoStop}%
\bibitem [{\citenamefont {Zhang}\ \emph {et~al.}(2015)\citenamefont {Zhang},
  \citenamefont {Martin},\ and\ \citenamefont
  {Newman}}]{zhang_identification_2015}%
  \BibitemOpen
  \bibfield  {author} {\bibinfo {author} {\bibfnamefont {Xiao}\ \bibnamefont
  {Zhang}}, \bibinfo {author} {\bibfnamefont {Travis}\ \bibnamefont {Martin}},
  \ and\ \bibinfo {author} {\bibfnamefont {M.~E.~J.}\ \bibnamefont {Newman}},\
  }\bibfield  {title} {\enquote {\bibinfo {title} {Identification of
  core-periphery structure in networks},}\ }\href {\doibase
  10.1103/PhysRevE.91.032803} {\bibfield  {journal} {\bibinfo  {journal}
  {Physical Review E}\ }\textbf {\bibinfo {volume} {91}},\ \bibinfo {pages}
  {032803} (\bibinfo {year} {2015})},\ \bibinfo {note} {publisher: American
  Physical Society}\BibitemShut {NoStop}%
\bibitem [{\citenamefont {Gallagher}\ \emph {et~al.}(2020)\citenamefont
  {Gallagher}, \citenamefont {Young},\ and\ \citenamefont
  {Welles}}]{gallagher_clarified_2020}%
  \BibitemOpen
  \bibfield  {author} {\bibinfo {author} {\bibfnamefont {Ryan~J.}\ \bibnamefont
  {Gallagher}}, \bibinfo {author} {\bibfnamefont {Jean-Gabriel}\ \bibnamefont
  {Young}}, \ and\ \bibinfo {author} {\bibfnamefont {Brooke~Foucault}\
  \bibnamefont {Welles}},\ }\bibfield  {title} {\enquote {\bibinfo {title} {A
  {Clarified} {Typology} of {Core}-{Periphery} {Structure} in {Networks}},}\
  }\href {http://arxiv.org/abs/2005.10191} {\bibfield  {journal} {\bibinfo
  {journal} {arXiv:2005.10191 [physics]}\ } (\bibinfo {year} {2020})},\
  \bibinfo {note} {arXiv: 2005.10191}\BibitemShut {NoStop}%
\bibitem [{\citenamefont {Boguñá}\ \emph {et~al.}(2010)\citenamefont
  {Boguñá}, \citenamefont {Papadopoulos},\ and\ \citenamefont
  {Krioukov}}]{boguna_sustaining_2010}%
  \BibitemOpen
  \bibfield  {author} {\bibinfo {author} {\bibfnamefont {Marián}\ \bibnamefont
  {Boguñá}}, \bibinfo {author} {\bibfnamefont {Fragkiskos}\ \bibnamefont
  {Papadopoulos}}, \ and\ \bibinfo {author} {\bibfnamefont {Dmitri}\
  \bibnamefont {Krioukov}},\ }\bibfield  {title} {\enquote {\bibinfo {title}
  {Sustaining the {Internet} with hyperbolic mapping},}\ }\href {\doibase
  10.1038/ncomms1063} {\bibfield  {journal} {\bibinfo  {journal} {Nat Commun}\
  }\textbf {\bibinfo {volume} {1}},\ \bibinfo {pages} {62} (\bibinfo {year}
  {2010})}\BibitemShut {NoStop}%
\bibitem [{\citenamefont {Newman}\ and\ \citenamefont
  {Peixoto}(2015)}]{newman_generalized_2015-1}%
  \BibitemOpen
  \bibfield  {author} {\bibinfo {author} {\bibfnamefont {M.~E.~J.}\
  \bibnamefont {Newman}}\ and\ \bibinfo {author} {\bibfnamefont {Tiago~P.}\
  \bibnamefont {Peixoto}},\ }\bibfield  {title} {\enquote {\bibinfo {title}
  {Generalized {Communities} in {Networks}},}\ }\href {\doibase
  10.1103/PhysRevLett.115.088701} {\bibfield  {journal} {\bibinfo  {journal}
  {Physical Review Letters}\ }\textbf {\bibinfo {volume} {115}},\ \bibinfo
  {pages} {088701} (\bibinfo {year} {2015})}\BibitemShut {NoStop}%
\end{thebibliography}%

\appendix
\section{Marginal likelihood of the non-uniform PP model}\label{app:nupp}
The model likelihood of the non-uniform PP model described in the text
can be written as
\begin{multline}
  P(\A|\bm\lambda,\omega,\bm\theta,\bb) =\\
  \frac{\ee^{-\omega\sum_{r<s}\hat\theta_r\hat\theta_s}\omega^{e_{\text{out}}}\prod_r\ee^{-\lambda_r\hat\theta_r^2/2}\lambda_r^{e_{rr}/2}\prod_i\theta_i^{k_i}}{\prod_{i<j}A_{ij}!\prod_iA_{ii}!!}.
\end{multline}
Enforcing the constraint $\hat\theta_r=1$, and using the noninformative
priors
\begin{align*}
  P(\lambda_r|\bar\lambda) &= \ee^{-\lambda_r/(2\bar\lambda)}/(2\bar\lambda)\\
  P(\omega|\bar\lambda) &= \ee^{-\omega/\bar\lambda}/\bar\lambda\\
  P(\theta|\bb) &= \prod_r(n_r-1)!\delta\left(\textstyle\sum_r\theta_i\delta_{b_i,r}-1\right),
\end{align*}
we obtain the following marginal likelihood, after integrating over all
parameters,
\begin{multline}
  P(\A|\lambda,\bb) =\\
  \frac{e_{\text{out}}!\prod_r(e_{rr}/2)!\prod_ik_i!}{\left(\frac{1}{2}+\frac{1}{2\bar\lambda}\right)^{e_{\text{in}}+B}\left[{B\choose 2}+\frac{1}{\lambda}\right]^{e_{\text{out}}+1}\prod_{i<j}A_{ij}!\prod_iA_{ii}!!}.
\end{multline}
This likelihood is once more identical to a microcanonical model,
\begin{multline}
  P(\A|\lambda,\bb) = P(\A|\bm e,\bm k,\bb) P(\bm k|\bm e,\bb)P(\bm e | \{e_{rr}\}, e_{\text{out}}, \bb) \\
  \times
  P(\{e_{rr}\} | \bar\lambda,\bb)
  P(e_{\text{out}}| \bar\lambda,\bb)
  P(E)
\end{multline}
where we now have the priors
\begin{align*}
  P(\bm e | \{e_{rr}\}, e_{\text{out}}, \bb) &= \frac{e_{\text{out}}!}{{B\choose 2}^{e_{\text{out}}}\prod_{r<s}e_{rs}!}\\
  P(\{e_{rr}\} | \bar\lambda,\bb) &= \prod_r\frac{\bar\lambda^{e_{rr}}}{(\bar\lambda + 1)^{e_{rr}+1}}\\
  P(e_{\text{out}}| \bar\lambda,\bb) &= \frac{\left[\bar\lambda{B\choose 2}\right]^{e_{\text{out}}}}{\left[\bar\lambda{B\choose 2}+1\right]^{e_{\text{out}}+1}}.
\end{align*}
We can improve this by replacing the last two equations with the following choice
\begin{align}
  P(\{e_{rr}\},e_{\text{out}}|\bb, E) &= P(\{e_{rr}\}|e_{\text{in}},\bb)P(e_{\text{in}}|E,\bb)\\
  &= {B + e_{\text{in}} - 1 \choose e_{\text{in}}}^{-1}
  \left(\frac{1}{E+1}\right)^{1-\delta_{B,1}},
\end{align}
which amounts to first choosing the value of $e_{\text{in}}$ uniformly
at random, and then likewise for the distribution of the diagonal values
$\{e_{rr}\}$. Multiplying the above equations as
\begin{multline}
  P(\A|\bb) = P(\A|\bm e,\bm k,\bb)  P(\bm k|\bm e,\bb)P(\bm e | \{e_{rr}\}, e_{\text{out}}, \bb)\times\\
  P(\{e_{rr}\},e_{\text{out}}|E, \bb) P(E)
\end{multline}
we arrive at
Eq.~\ref{eq:nupp} in the main text.

\section{General equivalences with statistical inference}\label{app:equiv}

Here we show that it not difficult to establish a formal connection
between any community detection method and statistical inference. Let us
consider an arbitrary quality function
\begin{equation}
  W(\A, \bb) \in \mathbb{R}
\end{equation}
that is used to perform community detection via the optimization
\begin{equation}
  \bb^* = \underset{\bb}{\operatorname{argmax}}\; W(\A, \bb).
\end{equation}
We can retrofit any such method, and transform it into a statistical
inference procedure by using $W(\A, \bb)$ as the Hamiltonian of an
\emph{ad hoc} generative model given by
\begin{equation}
  P(\A|\bb) = \frac{\ee^{W(\A,\bb)}}{Z(\bb)},
\end{equation}
with normalization given by
\begin{equation}
  Z(\bb) = \sum_{\A}\ee^{W(\A,\bb)}.
\end{equation}
In general, performing a maximum likelihood estimation of this model
will not be equivalent to the original optimization problem, due to the
role of the normalization constant $Z(\bb)$. However, we can cast it as
a Bayesian procedure in order to achieve a trivial equivalence, via the
posterior distribution
\begin{equation}
  P(\bb|\A) = \frac{P(\A|\bb)P(\bb)}{P(\A)},
\end{equation}
and by choosing the prior
\begin{equation}
  P(\bb) = \frac{Z(\bb)}{\Omega},
\end{equation}
with $\Omega = \sum_{\bb}Z(\bb)$, and
$P(\A)=\sum_{\bb}\ee^{W(A,\bb)}/\Omega$. Based on this, we recover easily
the original optimization by maximizing from the posterior distribution,
\begin{align}
  \bb^* &= \underset{\bb}{\operatorname{argmax}}\;\ln P(\bb|\A)\\
  & = \underset{\bb}{\operatorname{argmax}}\; W(\A, \bb).
\end{align}
Therefore, finding a mere equivalence between any given community
detection method and statistical inference, by itself, is not a very
insightful exercise, as it can amount to little more than tautology.
This also shows that not every inference procedure is any more
meaningful or principled than using an arbitrary quality
function. Instead, these features are contingent on the actual
generative models used, which need to be properly justified, together
with the choice of priors, and care should be taken to verify the
consistency of the whole approach, which is not granted automatically in
every case.

Despite the above, it should be mentioned that constructing a posterior
distribution in the \emph{ad hoc} way described above does have its
uses. In particular, it allows us to formally define a distribution over
all possible divisions of the network according to any given community
detection method. As shown in Ref.~\cite{zhang_scalable_2014}, by
characterizing this entire distribution, we have, to some extent, a
mechanism to detect degeneracy and evaluate the statistical significance
of the results, by seeking the consensus of a large fraction of the
solutions. Nonetheless, this does not address the arbitrariness of the
Hamiltonian chosen, and the ultimate interpretation of the results.

\end{document}